\documentstyle[prd,aps]{revtex}

\input epsfig.sty

\newcommand{\prepr}[1] {\begin{flushright}  {\bf #1} \end{flushright}
\vskip 1.cm}
\newcommand{\titul}[1] {\begin{center}{\Large {\bf #1 } } \end{center}
\vskip 0.8cm}

\newcommand{\autor}[1] {\begin{center}  {\bf \lineskip .3cm #1  }
                        \end{center} }

\newcommand{\lugar}[1] {\begin{center}  {\normalsize \bf \it #1   }
\end{center}}

\def\half{\frac{1}{2}}

\def\PR#1#2#3 {{\it Phys. Rev. }{\bf D#1} #2 {(#3)} }
\def\PRL#1#2#3 {{\it Phys. Rev. Lett. }{\bf #1} #2 {(#3)} }
\def\PL#1#2#3 {{\it Phys. Lett. }{\bf #1} #2 {(#3)}  }

\def\qbar{\overline q}

\def\be{\begin{equation}}
\def\ee{\end{equation}}
\def\ba{\begin{eqnarray}}
\def\ea{\end{eqnarray}}

\newcounter{muni}

\begin{document}
\hbadness=10000
\pagenumbering{arabic}
\begin{titlepage}
\prepr{
\hspace{20mm} NCKU-HEP-01-04 \\
\hspace{20mm} DPNU-00-13}
\titul{\bf
Leading-power contributions to $B\to\pi,\rho$ transition form factors}
\autor{T. Kurimoto$^1$\footnote{E-mail: krmt@sci.toyama-u.ac.jp},
Hsiang-nan Li$^{2}$ \footnote{E-mail: hnli@mail.ncku.edu.tw} and
A.I. Sanda$^3$\footnote{E-mail: sanda@eken.phys.nagoya-u.ac.jp}
}
\lugar{ $^{1}$ Department of Physics, Toyama University, Toyama 930-8555,
Japan}
\lugar{ $^{2}$ Department of Physics, National Cheng-Kung University,\\
Tainan, Taiwan 701, Republic of China}
\lugar{ $^{2}$ National Center for Theoretical Sciences,
Hsinchu, Taiwan 300, Republic of China}
\lugar{ $^{3}$ Department of Physics, Nagoya University, Nagoya 464-8602,
Japan}

\vskip 2.0cm
{\bf  PACS index : 13.25.Hw, 11.10.Hi, 12.38.Bx, 13.25.Ft}

\thispagestyle{empty}
\vspace{10mm}
\begin{abstract}

We calculate the $B\to \pi,~\rho$ transition form factors in the
framework of perturbative QCD to leading power of $1/M_B$, $M_B$ being
the $B$ meson mass. We explain the basic principle by discussing the
pion electromagnetic form factor. It is shown that the logarithmic and
linear singularities occurring at small momentum fractions of light
meson distribution amplitudes do not exist in a self-consistent
perturbative analysis, which includes $k_\perp$ and threshold resummations.

\end{abstract}
\thispagestyle{empty}
\end{titlepage}

\section{INTRODUCTION}

Branching ratios of $B$ meson two-body nonleptonic decays have been
measured by CLEOIII, Belle and Babar collaborations \cite{YK,a,b,c}.
CP violations in these modes may be observed in near future.
Cognizant of this point, we have presented some theoretical anticipations
for the $B\to K\pi$ \cite{KLS}, $\pi\pi,~\pi\rho$ \cite{LUY}, and $KK$
\cite{CL2} decays in the perturbative QCD (PQCD) framework. In particular,
$5\sim 15\%$ CP violation is expected in the $B\to K\pi$ decays. The
$B\to\pi,~\rho$ transition form factors are the integral part of
two-body nonleptonic decay amplitudes. In this paper we shall convince
readers that these form factors in the large recoil region of light
mesons are calculable in PQCD. This is where our approach starts
to differ from other approaches to exclusive $B$ meson decays.

According to PQCD factorization theorem, a form factor is written as
the convolution of a hard amplitude with initial-state and final-state
hadron distribution amplitudes $\phi(x)$, where $x$ is the momentum
fraction associated with one of the partons. It has been pointed out that
perturbative evaluation of the pion form factor suffers nonperturbative
enhancement from the end-point region with a momentum fraction $x\to 0$
\cite{IL}. If this is true, the hard amplitude is characterized by a low
scale, such that expansion in terms of a large coupling
constant $\alpha_s$ is not reliable. More serious end-point (logarithmic)
singularities have been observed in the twist-2 (leading-twist) contribution
to the $B\to\pi$ transition form factor \cite{SHB,ASY}. The singularities
even become linear at twist 3 (next-to-leading twist) \cite{BF}. Because
of these singularities, it was claimed that the $B\to\pi$ form factor is
dominated by soft dynamics and not calculable in PQCD \cite{KR}. We
shall argue that this conclusion is false. We shall show that at the end
points, where the above singularities occur, the double logarithms
$\alpha_s\ln^2 x$ should be resummed in order to justify perturbative
expansion. The result, called threshold resummation \cite{S0,CT}, leads to 
strong Sudakov suppression at $x\to 0$ \cite{L3}. Therefore, the end-point
singularities do not exist in a self-consistent PQCD analysis.

In this work we shall investigate contributions to the $B\to \pi$
and $B\to\rho$ transition form factors from twist-2 and from two-parton
twist-3 distribution amplitudes.

In Sec. II we illustrate the PQCD formalism by studying
the pion electromagnetic form factor. We review the
reasoning why one might conclude that the form factor is not calculable, 
and explain why these objections are not justified in QCD.

In Secs.~III and IV we derive the $B$ meson transition form factors.
It will be shown that the twist-3 contributions, which seem to be
proportional to $m_0/M_B$ or $M_\rho/M_B$, do not vanish in the
$M_B\to\infty$ limit. Here $m_0$, $M_\rho$, and $M_B$ are the chiral symmetry
breaking scale,  $\rho$ meson mass, and $B$ meson mass, respectively.
We record our results of the form factors at large recoil: the
$B\to\pi$ form factor $F_+\sim 0.3$ and the $B\to\rho$ form factor
$A_0\sim 0.4$.

Meson distribution amplitudes are defined and the Sudakov factor from
threshold resummation is derived in the Appendices.

\section{PQCD APPROACH TO FORM FACTORS}

The suggestion that a hadronic form factor is calculable in PQCD was
first made in Ref.~\cite{LB,BL,ER}. The rough idea is summarized as
follows. One expands the bound-state wave function for a pion in terms of
Fock states containing on-shell partons (quarks or gluons) \cite{LB},
\begin{eqnarray}
\Psi_M=\psi(q\qbar)+\psi(q\qbar g)+\psi(q\qbar gg)
+\psi(q\qbar q\qbar)+\psi(q\qbar q\qbar g)+\cdots  \;.
\end{eqnarray}
Define a soft function $\Psi_M(\Lambda)$ at a typical hadronic scale
$\Lambda$ as the initial wave function,
\begin{eqnarray}
\Psi_M(\Lambda)=\psi^\Lambda(q\qbar)
+\psi^\Lambda(q\qbar g)+\psi^\Lambda(q\qbar gg)
+\psi^\Lambda(q\qbar q\qbar)+\psi^\Lambda(q\qbar q\qbar g)+\cdots\;.
\end{eqnarray}
The wave function $\Psi_M$ can be related to $\Psi_M(\Lambda)$ via
\begin{eqnarray}
\Psi_M=\Psi_M(\Lambda)+G^\Lambda K\Psi_M(\Lambda)\;,
\end{eqnarray}
where $K$ is an irreducible kernel and $G^\Lambda$ the Green function
involving only hard loop momenta.

\begin{figure}
\begin{center}
\epsfig{file=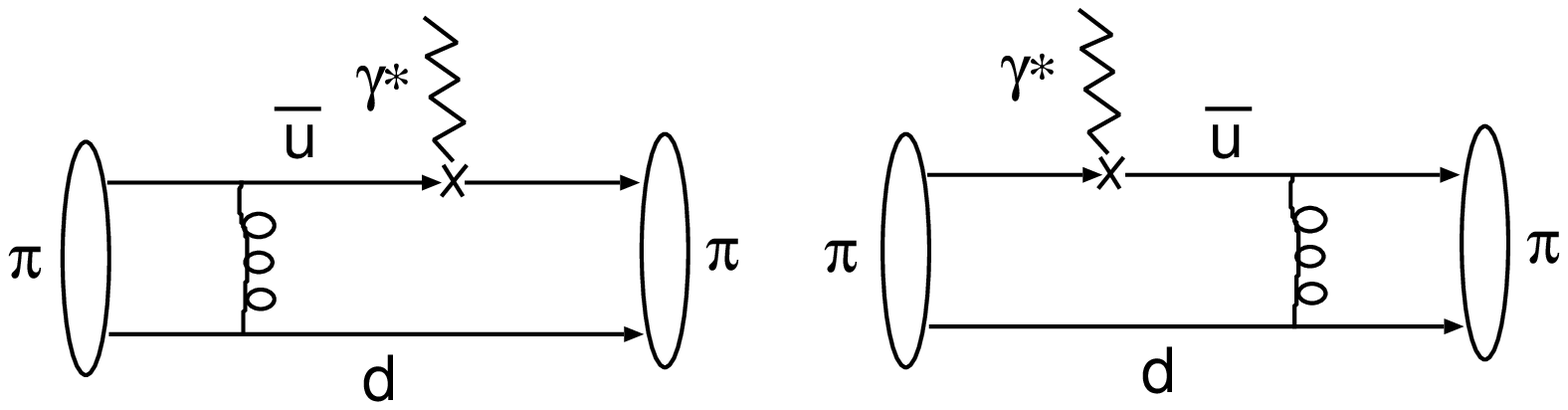,width=10cm}
\end{center}
\caption{Leading-order contribution to $F_\pi(Q^2)$.}
\label{leading}
\end{figure}

The pion electromagnetic form factor $F_\pi(Q^2)$ is then expressed as a
convolution integral,
\begin{eqnarray}
F_\pi(Q^2)=\int dx_1dx_2d^2k_{1\perp}d^2k_{2\perp}
\psi^\Lambda(P_1,x_1,\vec k_{1\perp})T_H(P_1,x_1,\vec k_{1\perp};
P_1+q,x_2,\vec k_{2\perp})
\psi^\Lambda(P_1+q,x_2,\vec k_{2\perp})+\cdots\;,
\label{fef}
\end{eqnarray}
with $P_1$ being the momentum of the initial-state pion, $q$ large
momentum transfer, and $Q^2=-q^2$. Here we have written the parton momenta 
associated with the initial state and final state as 
$k_1=(x_1Q/2,\vec k_{1\perp},x_1Q/2)$ and
$k_2=(x_2Q/2,\vec k_{2\perp},-x_2Q/2)$, respectively, in the notation
$p^\mu=(p^0,p^1,p^2,p^3)$, and made explicit the dependence of the two-parton
wave function $\psi^\Lambda(q\bar q)\equiv\psi^\Lambda(P_1,x_1,\vec k_{1\perp})$ 
on $P_1$ and $k_1$. The first term in Eq.~(\ref{fef}) contains
leading contributions, and ellipses represent those from higher Fock
states, which are down by powers of $1/Q^2$ in the light-cone gauge and
by powers of $\alpha_s$. The leading diagrams are displayed in
Fig.~\ref{leading}. It can be shown that the large momentum transfer
$Q^2$ flows through the hard amplitude $T_H$, and that all nonperturbative
dynamics goes into wave functions. One can therefore compute $T_H$
perturbatively.

However, it was pointed out that the above argument suffers a grave
difficulty \cite{IL}: the diagrams in Fig.~1 may be infrared divergent,
because important contribution to the form factor comes from the region,
where the exchanged gluons are soft. PQCD is then not applicable.
Below we shall examine this difficulty in more details. According to
Eq.~(\ref{fef}), the first diagram in Fig.~\ref{leading} gives
\begin{eqnarray}
\langle\pi(P_2)|J_\mu(0)|\pi(P_1)\rangle&=&
g^2C_F N_c\int dx_1dx_2d^2k_{1\perp}d^2k_{2\perp}
\frac{d z^-d^2 z_\perp}{(2\pi)^3}
\frac{dy^+d^2y_\perp}{(2\pi)^3}e^{-ik_2\cdot y}
\langle\pi(P_2)|{\bar d}_{\gamma}(y)u_{\beta}(0)|0\rangle
\nonumber\\
&& \times e^{ik_1\cdot z}\langle 0|{\bar u}_{\alpha}(0)d_{\delta}(z)|
\pi(P_1)\rangle~T_{H\mu}^{\gamma\beta;\alpha\delta}\;,
\label{pff}
\end{eqnarray}
with the color factor $C_F=4/3$, the number of colors $N_c=3$, and
the hard amplitude
\begin{eqnarray}
T_{H\mu}^{\gamma\beta;\alpha\delta}=[\gamma_\sigma]^{\gamma\delta}
\frac{1}{(k_2-k_1)^2}\left[\gamma^\sigma\frac{\not k_2-\not P_1}
{(P_1-k_2)^2}\gamma_\mu \right]^{\alpha\beta}\;.
\label{perp}
\end{eqnarray}
Write $(k_2-k_1)^2\sim -x_1x_2Q^2-|\vec k_{1\perp}-\vec k_{2\perp}|^2$.
If we ignore $|\vec k_{1\perp}-\vec k_{2\perp}|^2$, it has been shown
that the integral in Eq.~(\ref{pff}) is dominated by contributions from
the end-point regions with $x_1$, $x_2\to 0$. If the pion wave function
does not vanish at $x\to 0$, the integral will be even infrared divergent.
If we somehow regulate the infrared singularity by an appropriate choice
of the wave function, the running coupling constant $\alpha_s(x_1x_2Q^2)$
evaluated at the hard gluon momentum $x_1x_2Q^2$ is still too large to
make sense out of the perturbative expansion.

\subsection{Feynman's picture of a form factor}

\begin{figure}
\begin{center}
\epsfig{file=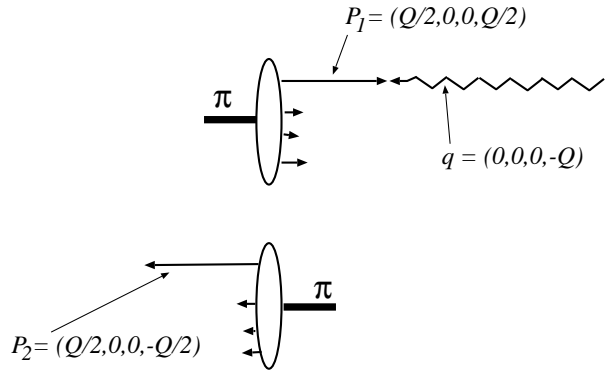,height=5cm}
\end{center}
\caption{Feynman's viewpoint of the dominant contribution to the pion
electromagnetic form factor.
}
\label{feynman}
\end{figure}

The above end-point singularity corresponds to the picture of the pion
form factor Feynman had in mind. In the so-called brick-wall frame the
initial-state pion with momentum $P_1=(Q/2,0,0,Q/2)$ is struck by a
space-like current of momentum $q=(0,0,0,-Q)$, and turns around with
momentum $P_2=(Q/2,0,0,-Q/2)$ as shown in Fig.~\ref{feynman}. Feynman
pointed out that the major contribution to the form factor comes from the
region, where one of the partons carries the full pion momentum. The rest
of partons, being all wee partons,
do not know in which direction they are moving. The resulting configuration
is essentially identical to the initial-state pion except that the momentum
of the fast parton is reversed. Hence, Feynman claimed that the $Q^2$
dependence of the pion form factor is related to the probability of
finding a single parton carrying all the pion momentum. Feynman's picture is
consistent with the statement that the form factor is dominated by the
singular part of Eq.~(\ref{perp}). Because it is singular, we can not
compute the form factor.

We argue that Feynman's picture of the pion form factor is false.
Consider a QED example. When an electron undergoes hard scattering,
it can not help but emit infinitely many photons in the direction of
the electron momentum. As a consequence, the elastic scattering cross 
section $\frac{d}{d\Omega}(e^+e^-\to e^+e^-)$ at finite angle vanishes 
at high energy, implying that the probability for the final $e^+e^-$ 
state being accompanied by no photons diminishes. In other words, the 
final state must be accompanied by many photons. In the QCD case of the 
pion form factor, when a quark inside the pion gets hit by a current, 
the final state will contain many gluons unless the spectator quark is 
nearby to shield the color charge. When one of the quarks carries all 
the momentum, the rest of the pion can not shield the color charge of 
the fast quark, and many gluons will be emitted in arbitrary directions 
during the hard scattering. Thus, the final configuration ending up as 
a single pion is extremely unlikely. This is so-called Sudakov 
suppression on exclusive processes at kinematic end points. Therefore, 
the contribution from Fig.~\ref{feynman} is negligible, and the 
end-point singularity does not exist!

\begin{figure}[h]
\begin{center}
\epsfig{file=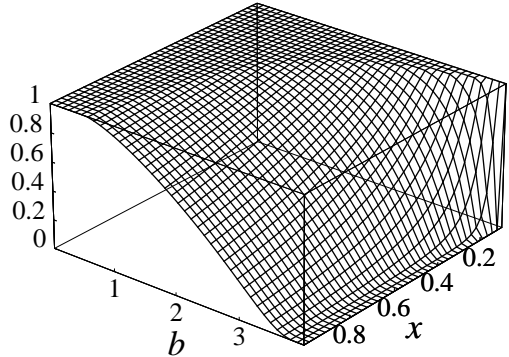,height=5cm}
\end{center}
\caption{The Sudakov factor $\exp[-S(x,b,P_1)]$. Note that its value is
very small in the region $b\sim b_{\rm max}=1/\Lambda_{\rm QCD}$,
with the QCD scale $\Lambda_{\rm QCD}=250$ MeV.}
\label{suda2}
\end{figure}

In the above argument we have ignored the term
$|\vec k_{1\perp}-\vec k_{2\perp}|^2$ in Eq.~(\ref{pff}). Small
$k_\perp$ in momentum space corresponds to large transverse distance
$b$ of two valence quarks. The color charge of the quark, which is
struck by the current, is not shielded in this large $b$ region, and will 
emit many gluons. The probability for having a single pion in the final
state is then vanishingly small. That is, this configuration can not 
contribute to the form factor. Hence, the momentum space with $k_\perp\to 0$, 
where the end-point singularity occurs, is also Sudakov
suppressed \cite{LS}. The typical behavior of the Sudakov factor
$\exp[-S(x,b,P_1)]$, $x=1-x_1$, which is associated with the struck quark,
is shown in Fig.~\ref{suda2}. We observe that the Sudakov factor decreases 
fast at large $b$ for $x\sim 1$ ($x_1\sim 0$), which
corresponds precisely to the end-point region in Eq.~(\ref{perp}). In
conclusion, the end-point singularity is absent, and the major
contribution to Fig.~\ref{leading} comes from the region with hard
gluon exchanges.

\subsection{Twist-3 contributions}

As derived in the Appendix A, a light-cone pion distribution amplitude
is written as
\begin{eqnarray}
\langle\pi^-(P)|{\bar d}_{\gamma}(y)u_{\beta}(0)|0\rangle
&=&-\frac{i}{\sqrt{2N_c}}
\int_0^1dx e^{ixP\cdot y}\left\{[\gamma_5\not P]_{\beta\gamma}\phi_\pi(x)
+ [\gamma_5]_{\beta\gamma}m_0\phi^p_\pi(x)\right.
\nonumber\\
& &\left.+m_0[\gamma_5(\not n_+\not n_--1)]_{\beta\gamma}
\phi^t_\pi(x)\right\}\;,
\label{fpd}
\end{eqnarray}
where $P=(P^+,0,0_\perp)$ is the pion momentum, the light-like vector 
$z=(0,z^-,0_\perp)$ the coordinate of the $d$ quark, and the dimensionless 
vector $n_+=(1,0,0_\perp)$ parallel to $P$ and $n_-=(0,1,0_\perp)$ 
parallel to $z$. Here a four-vector has been expressed in terms of 
light-cone coordinates,
\begin{eqnarray}
p^\mu=\left(\frac{p^0+p^3}{\sqrt{2}},\frac{p^0-p^3}{\sqrt{2}},
p_\perp\right)\;.
\end{eqnarray}
The distribution amplitude $\phi_\pi$ is twist-2, and $\phi_\pi^t$ and
$\phi_\pi^p$ proportional to $m_0=M_\pi^2/(m_d+m_u)\sim 1.4$ GeV, where
$m_q$ is the current quark mass of the quark $q$, are twist-3. The origin 
of these terms can be simply understood by means of the field-current 
identity from chiral symmetry,
\begin{eqnarray}
\overline d\gamma_5u=im_0f_\pi\pi^+\;.
\end{eqnarray}

It is easy to observe that twist-3 contributions are suppressed by a power
of $m_0/Q$. The asymptotic behaviors of $\phi_\pi$, $\phi_\pi^t$ and
$\phi_\pi^p$ are known to be
\begin{eqnarray}
\phi_\pi(x)\propto x(1-x)\;,\;\;\;
\phi^{p,t}_\pi(x)\propto 1\;.
\end{eqnarray}
As the hard amplitude in Eq.~(\ref{perp}) is convoluted with these
distribution amplitudes, we find that twist-2 contribution is finite,
while twist-3 ones are logarithmically divergent without Sudakov 
suppression. The Sudakov factor then introduces an effective cut-off to 
the integral at $x_c\sim \Lambda_{\rm QCD}/Q$, and the twist-3 
contributions are proportional to $(m_0/Q)\ln(Q/\Lambda_{\rm QCD})$. That 
is, the power counting is not altered by a logarithmic divergence in the
factorization formula. As shown later, the power counting for
contributions to the $B$ meson transition form factors is modified
by linear divergences in the factorization formulas. Therefore, the different 
end-point behavior leads to different power counting rules for the pion form 
factor and for the $B$ meson transition form factors.

\section{$B\to\pi$ transition form factors}

In the $B$ meson rest frame, we define the $B$ meson momentum $P_1$
and the pion momentum $P_2$ in the light-cone coordinates:
\begin{eqnarray}
P_1=\frac{M_B}{\sqrt{2}}(1,1,0_\perp)\;,\;\;\;
P_2=\frac{M_B}{\sqrt{2}}(\eta,0,0_\perp)\;,
\label{pa}
\end{eqnarray}
with the energy fraction $\eta$ carried by the pion. The spectator momenta
$k_1$ on the $B$ meson side and $k_2$ on the pion side are parametrized as
\begin{eqnarray}
k_1=\left(0,x_1\frac{M_B}{\sqrt{2}},{\vec k}_{1\perp}\right)\;,\;\;\;
k_2=\left(x_2\eta\frac{M_B}{\sqrt{2}},0,{\vec k}_{2\perp}\right)\;.
\label{k12}
\end{eqnarray}
Note that the four components of $k_1$ should be of the same order,
$O(\bar\Lambda)$, with $\bar\Lambda\equiv M_B-m_b$, $m_b$ being the $b$
quark mass. However, since $k_2$ is mainly in the plus direction
with $k_2^+\sim O(M_B)$, the hard amplitudes will not depend on the plus
component $k_1^+$ as explained below. This is the reason we do not
show $k_1^+$ in Eq.~(\ref{k12}) explicitly.

\begin{figure}[h]
\begin{center}
\epsfig{file=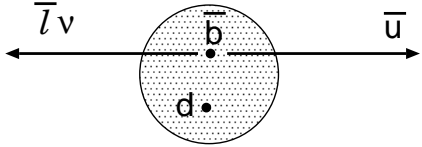,height=2cm}
\end{center}
\caption{Soft contribution to $F^{B\pi}$.}
\label{aa}
\end{figure}

Consider the configuration for the semileptonic decay
$B\to\pi \bar l\nu$ depicted in Fig.~\ref{aa}, which corresponds
to soft contribution to the $B\to\pi$ form factor $F^{B\pi}$. The $\bar u$
quark and the lepton pair fly back to back with energy of $M_B/2$.
The spectator quark $d$ carries a momentum of $O(\bar\Lambda)$.
If this configuration is responsible for the decay, it is impossible to
compute $F^{B\pi}$ using PQCD. However, applying an argument similar
to that used for the pion form factor, we know that the $\bar u$ quark 
recoiling against the lepton pair is bound to emit infinitely many gluons. 
Thus, Fig.~\ref{aa} in fact corresponds to the inclusive decay 
$B\to X_{\bar u}\bar l\nu$. The probability that the final state in 
Fig.~\ref{aa} contains only a single pion is suppressed by the
Sudakov form factors. A quantitative estimate of Sudakov suppression
of the soft contribution to $F^{B\pi}$ in the QCD sum rule formalism 
will be discussed later.

\subsection{Threshold and $k_\perp$ resummations}

It has been explained that the internal $\bar b$ quark involved in the hard
amplitude becomes on-shell as the momentum fraction $x$ of the $d$
quark vanishes \cite{L3}. The contributions to the $B\to \pi$ form factor
$F^{B\pi}$ are then logarithmically divergent at twist 2 and
linearly divergent at twist 3. We argue that as the end-point region
is important, the corresponding large double logarithms
$\alpha_s\ln^2 x$ need to be organized into a jet function $S_t(x)$ as
a consequence of threshold resummation \cite{L3}. This jet function
vanishes as $x\to 0, 1$, and modifies the end-point behavior of
meson distribution amplitudes effectively. This modification provides a
plausible explanation for the model of the twist-3 pion distribution
amplitude proportional to $x(1-x)$, which was adopted in \cite{KLS}. Our 
numerical study shows that the results of the $B\to\pi$ form factor 
obtained in this work are almost the same as those obtained in \cite{KLS}. 
In the following analysis we shall employ the approximate form,
\begin{eqnarray}
S_t(x)=\frac{2^{1+2c}\Gamma(3/2+c)}{\sqrt{\pi}\Gamma(1+c)} [x(1-x)]^c\;,
\label{trs}
\end{eqnarray}
where the parameter $c\approx 0.3$ comes from the best fit to the
next-to-leading-logarithm threshold resummation in moment space. Note
that the jet function $S_t$ is normalized to unity. For details of the
derivation, refer to the Appendix D.

Similarly, the inclusion of $k_\perp$ regulates the end-point
singularities, and large double logarithms $\alpha_s\ln^2 k_\perp$
are produced from higher-order corrections. These double logarithms
should be also organized to all orders, leading to $k_\perp$
resummation \cite{CS,BS}. The resultant Sudakov form factor, whose
explicit expression can be found in our previous works \cite{LY1,YL},
controls the magnitude of $k_\perp^2$ to be roughly
$O(\bar\Lambda M_B)$ by suppressing the region with
$k_\perp^2\sim O(\bar\Lambda^2)$. The coupling constant
$\alpha_s(\bar\Lambda M_B)/\pi \sim 0.13$ is then small enough to
justify the PQCD evaluation of heavy-to-light form factors \cite{KLS}.
We emphasize that the hard scale for heavy-to-light decays must be
$\bar\Lambda M_B$ in order to define a gauge-invariant $B$ meson 
distribution amplitude \cite{L4}. We shall include the Sudakov 
factor associated with the light spectator quark of the $B$ meson. Whether 
this factor is essential will be determined by the $B$ meson distribution 
amplitude. Since the $B$ meson is dominated by soft dynamics with 
$x_1\sim O(\bar\Lambda/M_B)$, the associated Sudakov effect is minor 
compared to that from the energetic pion.

With the possible order of magnitude of $k_\perp^2\sim O(\bar\Lambda M_B)$, 
a Taylor expansion of the hard gluon propagator near the end point,
\begin{eqnarray}
\frac{1}{(k_1-k_2)^2}\approx 
\frac{-1}{2k_1^-k_2^++|\vec k_{1\perp}-\vec k_{2\perp}|^2}\approx
\frac{-1}{x_1x_2\eta M_B^2}+
\frac{|\vec k_{1\perp}-\vec k_{2\perp}|^2}{(x_1x_2\eta M_B^2)^2}+\cdots
\label{fex}
\end{eqnarray}
is certainly not appropriate. A more reasonable treatment is to keep 
$k_\perp^2$ in the denominators of internal particle propagators, and to 
drop $k_\perp^2$ in the numerators, which are power-suppressed 
compared to other $O(M_B^2)$ terms. Under this prescription, the Sudakov
factor from $k_\perp$ resummation can be introduced into PQCD factorization
theorem without breaking gauge invariance of the hard amplitudes. For the 
same reason, the terms proportional to $k_1\sim O(\bar\Lambda)$ in the 
numerators should be neglected. It is then obvious from Eq.~(\ref{fex})
that the hard amplitudes are independent of the component $k_1^+$. The
$k_1^+$ dependence of the $B$ meson wave function can then be integrated 
out \cite{L4}, leading to the parametrization in Eq.~(\ref{k12}).

\begin{figure}[h]
\begin{center}
\epsfig{file=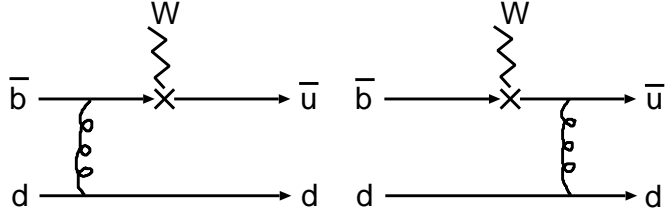,height=3cm}
\end{center}
\caption{Leading-order contribution to $F^{B\pi}$.}
\label{aa2}
\end{figure}

Note that the mechanism of threshold and $k_\perp$ resummations
is similar with the former responsible for suppression in the longitudinal
direction and the latter for suppression in the transverse direction.
As shown below, both twist-2 and twist-3 contributions are well-behaved
after including threshold and $k_\perp$ resummations. Hence, the
contributions to $F^{B\pi}$ from Fig.~\ref{aa2} dominate in the large
recoil region. In this configuration the $d$ quark gains a large
momentum parallel to the $\bar u$ quark momentum by exchanging a hard gluon
with the $\bar b$ or $\bar u$ quark.

\subsection{Form factors}

We compute the $B\to\pi$ form factors $F_+$ and $F_0$
defined by the following matrix element,
\begin{eqnarray}
\langle\pi(P_2)|{\bar b}(0)\gamma_\mu u(0)|B(P_1)\rangle
=F_+(q^2)\left[(P_1+P_2)_\mu-\frac{M_B^2-M_\pi^2}{q^2}q_\mu\right]
+F_0(q^2)\frac{M_B^2-M_\pi^2}{q^2}q_\mu\;,
\end{eqnarray}
where $q=P_1-P_2$ is the lepton-pair momentum. Another equivalent
definition is
\begin{eqnarray}
\langle\pi(P_2)|{\bar b}(0)\gamma_\mu u(0)|B(P_1)\rangle
=f_1(q^2)P_{1\mu}+f_2(q^2)P_{2\mu}\;,
\end{eqnarray}
in which the form factors $f_1$ and $f_2$ are related to $F_+$
and $F_0$ by
\begin{eqnarray}
F_+&=&\half(f_1+f_2)\;,
\\
F_0&=&\half f_1\left(1+ \frac{q^2}{M_B^2}\right)
+\half f_2\left(1 - \frac{q^2}{M_B^2}\right)\;.
\end{eqnarray}
The factorization formula for the $B\to\pi$ form factors is written as
\begin{eqnarray}
\langle\pi(P_2)|{\bar b}(0)\gamma_\mu u(0)|B(P_1)\rangle&=&
g^2C_F N_c\int dx_1 dx_2d^2k_{1\perp}d^2k_{2\perp}
\frac{dz^+d^2z_\perp}{(2\pi)^3}\frac{dy^-d^2y_\perp}{(2\pi)^3}
e^{-ik_2\cdot y}\langle\pi(P_2)|{\bar d}_{\gamma}(y)u_{\beta}(0)|0\rangle
\nonumber \\
& & \times e^{ik_1\cdot z}\langle 0|{\bar b}_{\alpha}(0)d_{\delta}(z)|
B(P_1)\rangle~T_{H\mu}^{\gamma\beta;\alpha\delta}\;.
\label{fbpi}
\end{eqnarray}
The pion distribution amplitude
$\langle\pi|{\bar d}_{\gamma}(y)u_{\beta}(0)|0\rangle$
has been supplied in Eq.~(\ref{fpd}), and the $B$ meson wave function
is given by (see the Appendix C)
\begin{eqnarray}
\int \frac{dz^+d^2z_\perp}{(2\pi)^3}e^{ik_1\cdot z}
\langle 0|{\bar b}_{\alpha}(0)d_{\delta}(z)|B(P_1)\rangle
=-\frac{i}{\sqrt{2N_c}}[(\not P_1+M_B)\gamma_5
\phi_B(k_1)]_{\alpha\delta}\;.
\label{bwp1}
\end{eqnarray}

Employing Eqs.~(\ref{fpd}) and (\ref{bwp1}), we derive, 
from Eq.~(\ref{fbpi}),
\begin{eqnarray}
f_1&=&16\pi M_B^2C_Fr_\pi\int dx_1dx_2\int b_1db_1 b_2db_2\phi_B(x_1,b_1)
[\phi_\pi^p(x_2)-\phi_\pi^t(x_2)] E(t^{(1)})h(x_1,x_2,b_1,b_2)\;,
\label{f1}\\
f_2&=&16\pi M_B^2C_F\int dx_1dx_2\int b_1db_1 b_2db_2\phi_B(x_1,b_1)
\nonumber\\
&\times& \Biggl\{\left[\phi_\pi(x_2)(1+x_2\eta)
+2r_\pi\left((\frac{1}{\eta} -x_2 )\phi_\pi^t(x_2) -x_2\phi_\pi^p(x_2)
 \right)\right]E(t^{(1)})h(x_1,x_2,b_1,b_2)
\nonumber\\
& &+ 2r_\pi\phi_\pi^p E(t^{(2})h(x_2,x_1,b_2,b_1)\Biggr\}\;,
\label{fpi}
\end{eqnarray}
with the ratio $r_\pi=m_0/M_B$ and the evolution factor
\begin{eqnarray}
E(t)=\alpha_s(t)e^{-S_B(t)-S_\pi(t)}\;.
\label{evol}
\end{eqnarray}
In the above formulas we have dropped the terms proportional to
the momentum fraction $x_1\sim O(\bar\Lambda/M_B)$ as argued before,
which are power-suppressed compared to the leading terms such as
$1+x_2/\eta$ in the form factor $f_2$.
The explicit expressions of the Sudakov exponents $S_B$ and $S_\pi$
are referred to \cite{LY1}. The hard function is written as
\begin{eqnarray}
h(x_1,x_2,b_1,b_2)&=&S_t(x_2)K_{0}\left(\sqrt{x_1x_2\eta}M_Bb_1\right)
\nonumber \\
& &\times \left[\theta(b_1-b_2)K_0\left(\sqrt{x_2\eta}M_B
b_1\right)I_0\left(\sqrt{x_2\eta}M_Bb_3\right)\right.
\nonumber \\
& &\left.+\theta(b_2-b_1)K_0\left(\sqrt{x_2\eta}M_Bb_2\right)
I_0\left(\sqrt{x_2\eta}M_Bb_1\right)\right]\;,
\label{dh}
\end{eqnarray}
where the factor $S_t$ suppresses the end-point behaviors of the pion
distribution amplitudes, especially of the twist-3 ones.
The hard scales $t$ are defined as
\begin{eqnarray}
t^{(1)}&=&{\rm max}(\sqrt{x_2\eta}M_B,1/b_1,1/b_2)\;,
\nonumber\\
t^{(2)}&=&{\rm max}(\sqrt{x_1\eta}M_B,1/b_1,1/b_2)\;.
\end{eqnarray}
It is obvious that if turning off threshold and $k_\perp$ resummations with
$\alpha_s$ fixed, Eqs.~(\ref{f1}) and (\ref{fpi}) are infrared divergent.

We argue that the two-parton twist-3 distribution amplitudes 
$\phi_\pi^{p,t}$, though proportional to the ratio $m_0/M_B$, need to be 
taken into account. As stated above, the corresponding convolution 
integrals for the $B\to\pi$ form factor are linearly divergent without
including Sudakov effects. These 
integrals, regulated in some way with an effective cut-off
$x_c\sim \bar\Lambda/M_B$, are proportional to the ratio
$M_B/\bar\Lambda$. Combining the two ratios $m_0/M_B$ and 
$M_B/\bar\Lambda$, the twist-3 contributions are in fact not down
by a power of $1/M_B$:
\begin{eqnarray}
\frac{m_0}{M_B}\int_{x_c}^1\frac{dx_2}{x_2^2}\sim
O\left(\frac{m_0}{\bar\Lambda}\right)\;,
\label{li}
\end{eqnarray}
and should be included in a complete leading-power
analysis. We emphasize that the presence of linear divergences
modifies the power counting rules, making the difference
between the $B$ meson transition form factors and the pion form
factor.

Various computing methods have been proposed for the evaluation of
the $B\to\pi$ transition form factors $F^{B\pi}(q^2)$ in the literature,
such as the lattice technique \cite{aoki}, light-cone QCD sum rules 
\cite{KR,PB3}, and PQCD \cite{LY1,DJK}. Obviously, lattice calculations 
become more difficult in the large recoil region of the light meson. 
However, this region is the one where PQCD is reliable, indicating that 
the PQCD and lattice approaches complement each other. This complementation 
will be explicitly exhibited in Fig.~\ref{bpi} below. In light-cone sum 
rules, dynamics of the $B\to\pi$ form factors have been assumed to be 
dominated by the large scale of $O(m_b)$. This is the reason twist 
expansion into Fock states in powers of $1/m_b$ applies to the pion bound 
state. If this assumption is valid, PQCD should be also applicable to the 
$B\to\pi$ form factors. Besides, large radiative correction to
the $B$ meson vertex, which reaches 35\% of the full contribution,
or about half of the soft (zeroth-order) contribution, has been noticed. 
This $O(\alpha_s)$ correction renders the sum rule for $f_BF^{B\pi}$,
with $f_B$ being the $B$ meson decay constant, quite unstable relative 
to the variation of input parameters \cite{PB3,KR2}. To stabilize the sum 
rule, one considers another sum rule for $f_B$ at the same time, which 
also receives large radiative correction to the $B$ meson vertex. The 
two large vertex corrections then cancel in the ratio $f_BF^{B\pi}/f_B$.
However, the radiative correction to $f_B$ is then large.

A careful look at the light-cone-sum-rule analyses indicates that the soft 
contribution is more sensitive to the end-point ($x\to 0$) behavior of 
the pion distribution amplitude than the $O(\alpha_s)$ correction 
\cite{KR2}. Hence, if the end-point behavior of the pion distribution 
amplitude is modified by the Sudakov factor in this work, such that the
end-point contribution is not important, perturbative contribution can 
become dominant. The Sudakov effect on the soft contribution to
$F^{B\pi}(0)$ has been investigated in the QCD sum rule formalism \cite{SR}
(without twist expansion for the pion bound state). In this analysis,
the soft contribution without Sudakov suppression was estimated to be
between 0.15 (corresponding to $f_B\sim 190$ MeV) and 0.22
(corresponding to $f_B\sim 130$ MeV). The soft contribution to
$f_BF^{B\pi}$ obtained in \cite{KR2} is consistent with the above
range. It was then shown that the Sudakov effect decreases the soft
contribution by a factor 0.4-0.7, depending on infrared cut-offs for
loop corrections to the weak decay vertex. Therefore, the soft
contribution turns out to be about 0.06-0.15. Compared with the
lattice results $F^{B\pi}(0)\sim 0.3$, it is reasonable to conclude
that the soft contribution amounts to about 30\%, which is consistent with the
observation made in \cite{LY1}. It is a fair opinion that the estimate of 
soft contribution is more model-dependent than perturbative one. For example, 
perturbative contribution is less sensitive to the pion distribution 
amplitude or to other input parameters such as the Borel mass in light-cone
sum rules \cite{KR2}. In the PQCD approach we calculate the perturbative 
contribution to $F^{B\pi}$, which is more model-independent, and show that 
the result can more or less saturate the value predicted by lattice technique.

For the $B$ meson distribution amplitude, we adopt the model
\begin{eqnarray}
\phi_B(x,b)=N_Bx^2(1-x)^2
\exp\left[-\frac{1}{2}\left(\frac{xM_B}{\omega_B}\right)^2
-\frac{\omega_B^2 b^2}{2}\right]\;,
\label{os}
\end{eqnarray}
with the shape parameter $\omega_B=0.4$ GeV \cite{KLS}. The
normalization constant $N_B$ is related to the decay constant
$f_B=190$ MeV through the relation
\begin{eqnarray}
\int dx_{1}\phi_B(x_{1},0)=\frac{f_B}{2\sqrt{2N_c}}\;.
\end{eqnarray}
It is easy to find that Eq.~(\ref{os}) has a maximum at 
$x\sim \bar\Lambda/M_B$. We employ the models for the pion \cite{PB2},
\begin{eqnarray}
\phi_\pi(x)&=&\frac{3f_\pi}{\sqrt{2N_c}} x(1-x)
\left[1+0.44C_2^{3/2}(2x-1)+0.25C_4^{3/2}(2x-1)\right]\;,
\label{pioa}\\
\phi_\pi^p(x)&=&\frac{f_\pi}{2\sqrt{2N_c}}
\left[1+0.43C_2^{1/2}(2x-1)+0.09C_4^{1/2}(2x-1)\right]\;,
\label{piob}\\
\phi_\pi^t(x)&=&\frac{f_\pi}{2\sqrt{2N_c}} (1-2x)
\left[1+0.55(10x^2-10x+1)\right]\;,
\label{pioc}
\end{eqnarray}
with the pion decay constant $f_\pi=130$ MeV. The Gegenbauer 
polynomials are defined by
\begin{eqnarray}
& &C_2^{1/2}(t)=\frac{1}{2}(3t^2-1)\;,\;\;\;
C_4^{1/2}(t)=\frac{1}{8}(35 t^4 -30 t^2 +3)\;,
\nonumber\\
& &C_2^{3/2}(t)=\frac{3}{2}(5t^2-1)\;,\;\;\;
C_4^{3/2}(t)=\frac{15}{8}(21 t^4 -14 t^2 +1) \;,
\end{eqnarray}
whose coefficients correspond to $m_0=1.4$ GeV.

We first investigate the relative importance of the twist-2 and 
twist-3 contributions to $F_+(q^2)$, and the results are listed in
Table~\ref{tbl-twist}. It is observed that the latter are in fact
larger than the former, consistent with the argument that the twist-3 
contributions are not power-suppressed. The light-cone sum rules also 
give approximately equal weights to the twist-2 and higher-twist 
contributions to $F_+$ \cite{KR2}. We then compare our results of 
$F_+(q^2)$ and $F_0(q^2)$ for $q^2 = 0 \sim 10$ GeV$^2$ with those 
derived from lattice QCD \cite{UKQCD} and from light-cone sum rules 
\cite{PB3} in Fig.~\ref{bpi}, where lattice results have been 
extrapolated to the small $q^2$ region. Different extrapolation 
methods cause uncertainty only of about 5\% \cite{DB}.  
The good agreement among these 
different approaches at large recoil is explicit. The fast rise of the
PQCD results at slow recoil indicates that perturbative calculation
becomes unreliable gradually. The values of $F_+(0)= F_0(0)\equiv F(0)$ 
from PQCD for the parameter $\omega_B = 0.40 \pm 0.04$ GeV are listed in
Table~\ref{Bpitable}. The resultant range $F_+(0)=0.30\pm 0.04$ is 
in agreement with $F(0)\sim 0.3$ obtained in \cite{PB3,UKQCD}. We shall
adopt the same range of $\omega_B$ in the evaluation of the $B\to\rho$
transition form factors below. We also examine the uncertainty of our 
predictions from the parametrization of the jet function in Eq.~(\ref{trs}).
The values of $F_+(q^2)$ vary about 15\% for the choices of $c=0.2$
and $c=0.4$ as shown in Table~\ref{rsumtbl-pls}.
The variation for $F_0(q^2)$ is similar.
In a future work we shall incorporate the exact jet function into a
convolution integrand in moment space.

\begin{table}
\begin{center}
\begin{tabular}{c||ccccccccccc}
$q^2$ (GeV${}^2$)&0.0&1.0&2.0&3.0&4.0&5.0&6.0&7.0&8.0&9.0&10.0\\\hline
twist 2 &0.120&0.128&0.138&0.148&0.159&0.172&0.188&0.204&0.223&0.243&0.270\\
twist 3 &0.177&0.193&0.210&0.230&0.253&0.279&0.308&0.344&0.385&0.432&0.487\\\hline
total&0.297&0.321&0.348&0.378&0.412&0.451&0.496&0.548&0.608&0.675&0.757\\
\end{tabular}
\end{center}
\caption{%
Contributions to $F_+(q^2)$ from the twist-2 and two-parton twist-3
pion distribution amplitudes.
}
\label{tbl-twist}
\end{table}

\begin{table}[t]
\begin{center}
\begin{tabular}{c||ccccccccc}
$\omega_B$ (GeV)\ \  & 0.36 & 0.37 & 0.38 & 0.39 & 0.40 & 0.41 & 0.42
& 0.43 & 0.44 \\ \hline
$F(0)$ \ \ & 0.345 & 0.334 & 0.321 & 0.309 & 0.297 & 0.287
& 0.277 & 0.268 & 0.259
\end{tabular}
\end{center}
\caption{%
Values of $F_+(0) = F_0(0)\equiv F(0)$ for given $\omega_B$.
}
\label{Bpitable}
\end{table}

\begin{table}
\begin{center}
\begin{tabular}{c||ccccccccccc}
$q^2$ (GeV${}^2$)&0.0&1.0&2.0&3.0&4.0&5.0&6.0&7.0&8.0&9.0&10.0\\\hline
$c=0.2$&0.347&0.376&0.406&0.442&0.482&0.527&0.580&0.639&0.709&0.790&0.886\\
$c=0.3$&0.297&0.321&0.348&0.378&0.412&0.451&0.496&0.548&0.608&0.675&0.757\\
$c=0.4$&0.260&0.280&0.303&0.330&0.359&0.392&0.432&0.475&0.527&0.588&0.659
\end{tabular}
\end{center}
\caption{%
Values of $F_+(q^2)$ for $c= 0.2$, 0.3, and 0.4.
}
\label{rsumtbl-pls}
\end{table}


\begin{figure}[b]
\begin{center}
\epsfig{file=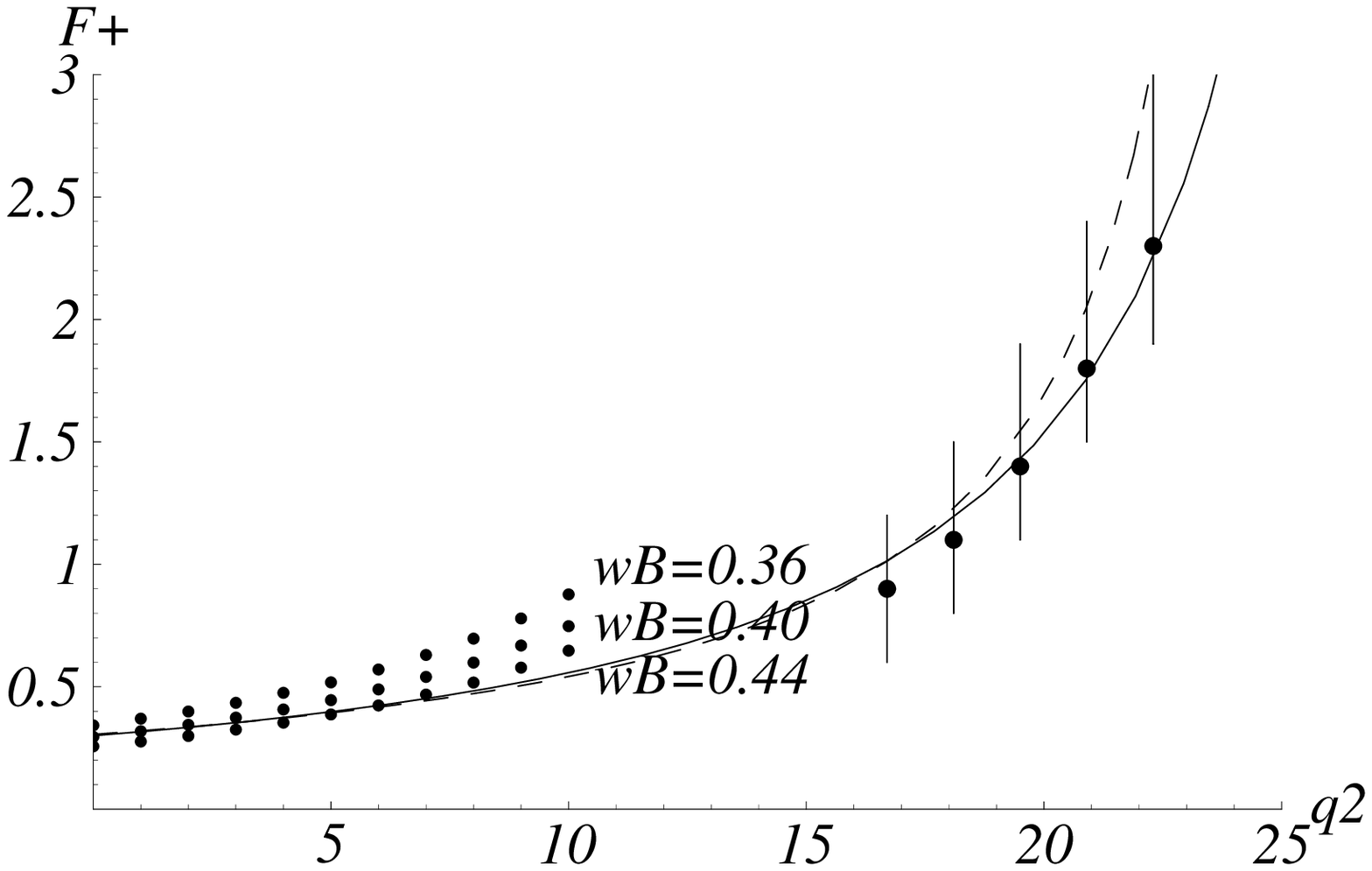,width=10cm}
\epsfig{file=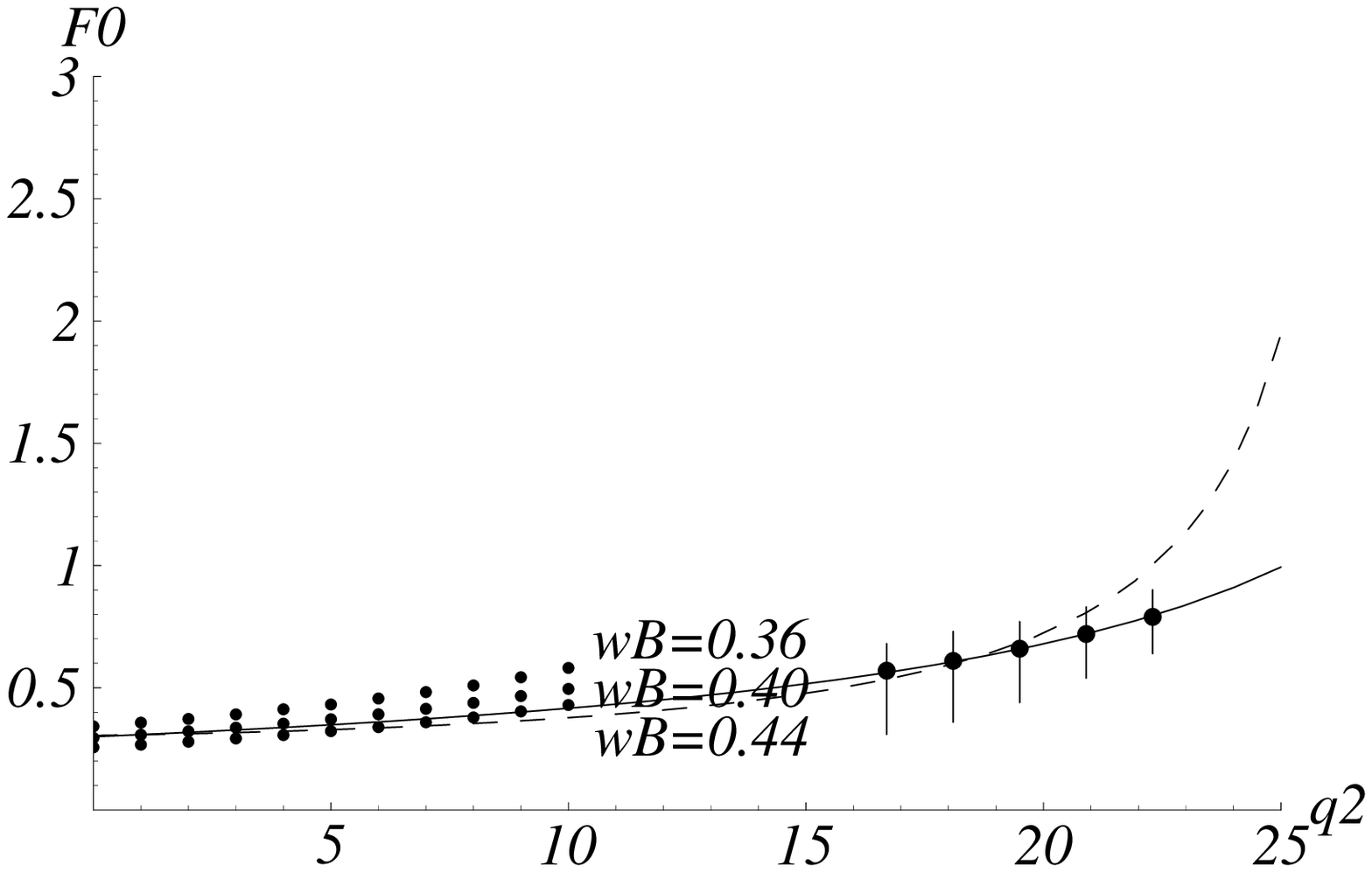,width=10cm}
\end{center}
\caption{
The $B\to \pi$ form factors $F_+$ and $F_0$ as
functions of $q^2$ (GeV$^2$). PQCD results for $\omega_B=0.36$,
0.40, and 0.44 GeV are shown in dots.
The solid lines correspond to fits to the lattice QCD results
with errors. The dashed lines come from light-cone sum rules.
}
\label{bpi}
\end{figure}

\section{$B\to\rho$ TRANSITION FORM FACTORS}

Consider the semileptonic decay $B\to\rho\bar l \nu$ in the fast
recoil region of the $\rho$ meson \cite{LY2}. We define the $B$ meson
momentum $P_1$ as in Eq.~(\ref{pa}), the momentum $P_2$ and the
polarization vectors $\epsilon$ of the $\rho$ meson in light-cone
coordinates as
\begin{eqnarray}
& &P_2=\frac{M_B}{\sqrt{2}\eta}(\eta^2,r_\rho^2,0_\perp)\;.
\nonumber\\
& &\epsilon_L=\frac{1}{\sqrt{2}r_\rho\eta}(\eta^2,-r_\rho^2, 0_\perp)\;,
\;\;\;\epsilon_T=(0,0,1,0) \mbox{ or } (0,0,0,1)\;,
\label{par}
\end{eqnarray}
with the ratio $r_\rho=M_\rho/M_B$ and the energy fraction $\eta$
carried by the $\rho$ meson. We first keep the $r^2_\rho$ dependence of
the kinematic variables in Eq.~(\ref{par}), and extract the twist-3 terms
proportional to $r_\rho$. The parametrization of $P_2$ and $\epsilon$
is chosen to make this extraction straightforward.

The $B\to\rho$ form factors are defined through the following
decompositions of hadronic matrix elements,
\begin{eqnarray}
\langle\rho(P_2,\epsilon^*)|{\bar b}(0)\gamma^\mu u(0)|B(P_1)\rangle
&=&\frac{2iV(q^2)}{M_B+M_\rho}\epsilon^{\mu\nu\rho\sigma}
\epsilon^*_\nu P_{2\rho} P_{1\sigma}\;,
\label{vf}\\
\langle\rho(P_2,\epsilon^*)|{\bar b}(0)\gamma^\mu\gamma_5 u(0)
|B(P_1)\rangle
&=&2M_\rho A_0(q^2)\frac{\epsilon^*\cdot q}{q^2}q^\mu+
(M_B+M_\rho)A_1(q^2)\left[\epsilon^{*\mu}-
\frac{\epsilon^*\cdot q}{q^2}q^\mu\right]
\nonumber\\
& &-A_2(q^2)\frac{\epsilon^*\cdot q}{M_B+M_\rho}
\left[P_1^\mu+P_2^\mu-\frac{M_B^2-M_\rho^2}{q^2}q^\mu\right]\;.
\label{af}
\end{eqnarray}
To calculate the form factors $V$, $A_0$ $A_1$ and $A_2$, we adopt the
following procedures. First, only the transverse polarization vectors
$\epsilon_T$ are involved in Eq.~(\ref{vf}) and associated with the
definition of $A_1$ in Eq.~(\ref{af}), through which we evaluate the form
factors $V$ and $A_1$, respectively. Both the structures
associated with $A_1$ and $A_2$ are orthogonal to the lepton pair momentum
$q$. Contracting Eq.~(\ref{af}) with $q_\mu$, we have
\begin{eqnarray}
\langle\rho(P_2,\epsilon^*)|{\bar b}(0)\not q\gamma_5 u(0)
|B(P_1)\rangle
=2M_\rho A_0(q^2)\epsilon^*\cdot q\;,
\label{a0e}
\end{eqnarray}
which implies that only the form factor $A_0$ is relevant in two-body
nonleptonic decays such as $B\to\rho \pi (K)$. We calculate $A_0$ from
Eq.~(\ref{a0e}) using the distribution amplitudes associated with a
longitudinally polarized $\rho$ meson.

For the longitudinal polarization vector $\epsilon_L$, the structures of 
$A_1$ and $A_2$ are in fact proportional to each other:
\begin{eqnarray}
\frac{\epsilon^*\cdot q}{M_B+M_\rho}
\left[P_1^\mu+P_2^\mu-\frac{M_B^2-M_\rho^2}{q^2}q^\mu\right]
=\frac{(\epsilon^*\cdot q)^2(M_B^2-M_\rho^2-q^2)}
{(M_B+M_\rho)[(\epsilon^*\cdot q)^2+q^2]}
\left[\epsilon^{*\mu}-\frac{\epsilon^*\cdot q}{q^2}q^\mu\right]\;,
\end{eqnarray}
which can be easily derived via the relation,
\begin{eqnarray}
P_1 +P_2=\frac{M_B^2-M_\rho^2-q^2}{(\epsilon^*\cdot q)^2+q^2}
\epsilon^*\cdot q\epsilon^*_L
+\frac{M_B^2-M_\rho^2+(\epsilon^*\cdot q)^2}{(\epsilon^*\cdot q)^2+q^2}q\;.
\end{eqnarray}
Contracting Eq.~(\ref{af}) with
$\epsilon_{*\mu}-\epsilon^*\cdot q q_\mu/q^2$, we obtain
\begin{eqnarray}
& &\langle\rho(P_2,\epsilon^*)|{\bar b}(0)
\left[\not\epsilon^*-\frac{\epsilon^*\cdot q}{q^2}\not q\right]
\gamma_5 u(0)|B(P_1)\rangle
\nonumber\\
& &=\frac{2P_2\cdot q}{M_B+M_\rho}\frac{(\epsilon^*\cdot q)^2}{q^2}
\left[A_2-\frac{(M_B+M_\rho)^2}{2P_2\cdot q}
\left(1+\frac{q^2}{(\epsilon^*\cdot q)^2}\right)A_1\right]\;,
\label{ea2}
\end{eqnarray}
from which the form factor $A_2$ can be computed. It turns out that
$A_1$ and $A_2$ have a simple relation, since the left-hand side of 
Eq.~(\ref{ea2}) is power-suppressed.

We derive the leading-power factorization formulas,
\begin{eqnarray}
V&=&8\pi M_B^2C_F\int dx_1dx_2\int b_1db_1 b_2db_2
\phi_B(x_1,b_1)
\nonumber\\
& &\times \Biggl\{\left[\phi_\rho^T(x_2)+r_\rho \left(
\left(\frac{2}{\eta}+x_2\right)\phi_\rho^a(x_2)-x_2\phi_\rho^v(x_2)
\right)\right]E(t^{(1)})h(x_1,x_2,b_1,b_2)
\nonumber \\
&&+ r_\rho\left[\phi_\rho^v(x_2)+\phi_\rho^a(x_2)\right]
E(t^{(2)})h(x_2,x_1,b_2,b_1)\Biggr\}\;,
\\
A_0&=&8\pi M_B^2 C_F\int dx_1dx_2\int b_1db_1
b_2db_2\phi_B(x_1,b_1)
\nonumber\\
& &\times\Biggl\{\left[(1+\eta x_2)\phi_\rho(x_2)+r_\rho
\left((1-2x_2)\phi_\rho^t(x_2)+
\left(\frac{2}{\eta}-1-2x_2\right)\phi_\rho^s(x_2)\right)\right]
E(t^{(1)})h(x_1,x_2,b_1,b_2)
\nonumber \\
& &+2r_\rho\phi_\rho^s(x_2)E(t^{(2)})h(x_2,x_1,b_2,b_1)\Biggr\}\;,
\\
A_1&=&8\pi M_B^2 C_F\eta\int dx_1dx_2\int b_1db_1
b_2db_2\phi_B(x_1,b_1)
\nonumber\\
& &\times\Biggl\{\left[\phi_\rho^T(x_2)+r_\rho
\left(\left(\frac{2}{\eta}+x_2\right)\phi_\rho^v(x_2)
-x_2\phi_\rho^a(x_2)\right)\right]E(t^{(1)})h(x_1,x_2,b_1,b_2)
\nonumber \\
& &+r_\rho\left[\phi_\rho^v(x_2)+\phi_\rho^a(x_2)\right]
E(t^{(2)})h(x_2,x_1,b_2,b_1)\Biggr\}\;,
\\
A_2&=&\frac{A_1}{\eta}\;,
\label{a2f}
\end{eqnarray}
with the evolution factor $E(t)$ the same as in Eq.~(\ref{evol}).
Taking the fast recoil limit with $\eta\to 1$ and assuming the asymptotic
behavior $\phi_\rho^v=\phi_\rho^a$, the above form factors 
are found to obey the symmetry relations \cite{BF,JC},
\begin{eqnarray}
V=A_1\;,\;\;\; A_2=A_1-2r_\rho A_0\;,
\label{sr}
\end{eqnarray}
where the term $-2r_\rho A_0$, being higher-power, does not appear
in Eq.~(\ref{a2f}). Note that the form factors, treated as 
nonperturbative objects, are not calculated in \cite{BF}. Instead, the 
diagrams we have calculated above are regarded as perturbative corrections 
to the relations in Eq.~(\ref{sr}).

\begin{figure}[h]
\begin{center}
\epsfig{file=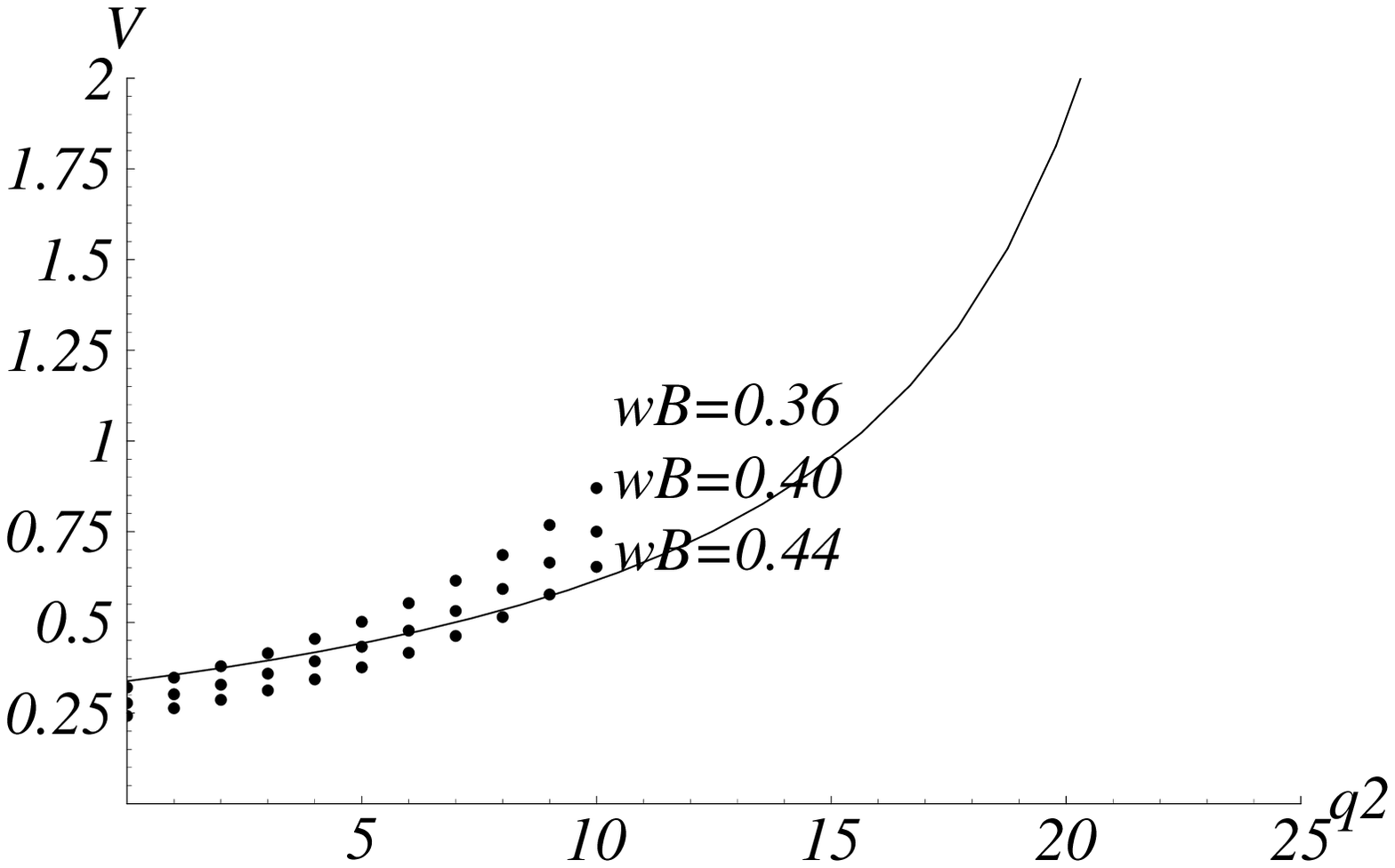,height=5.3cm}
\epsfig{file=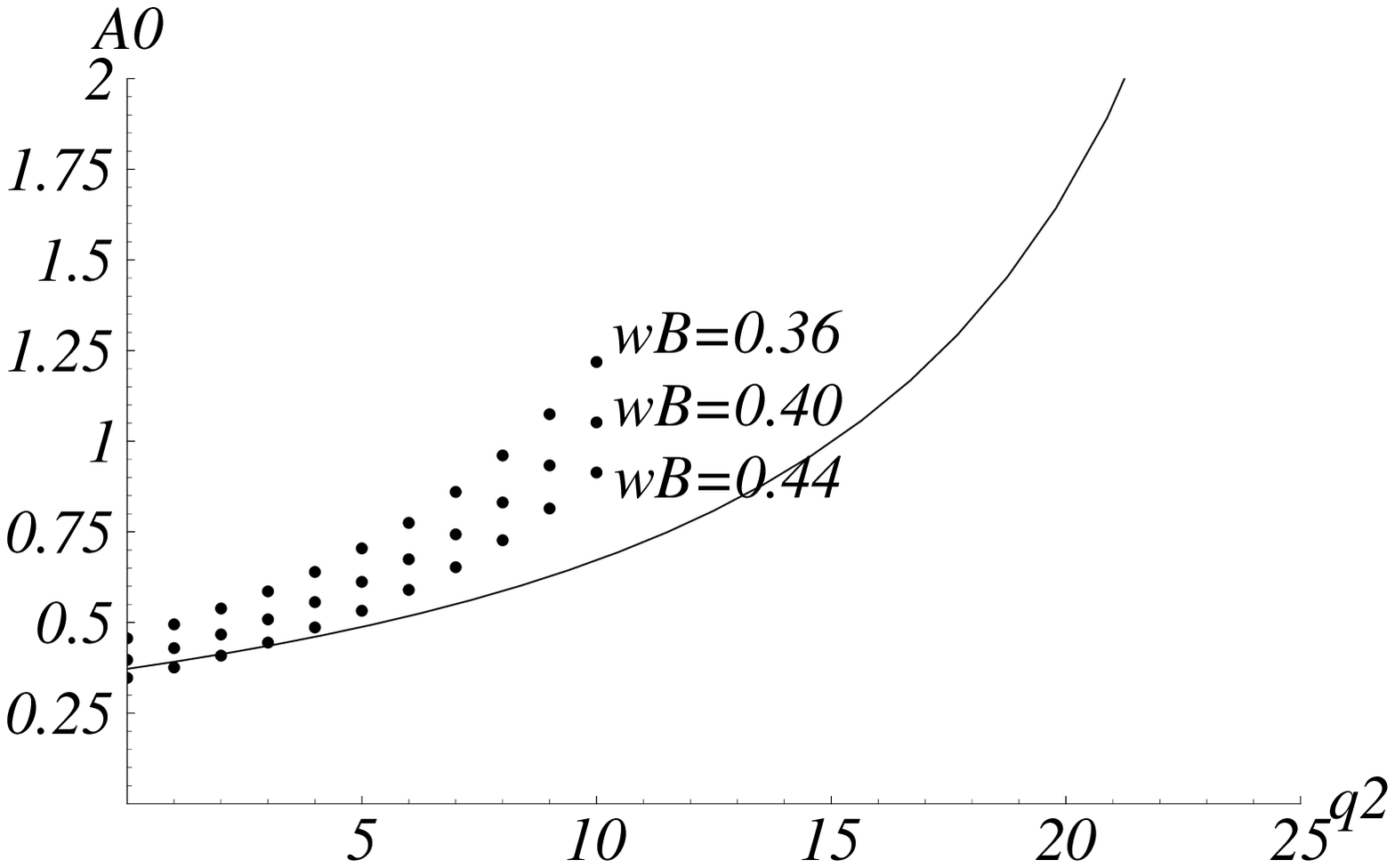,height=5.3cm}
\epsfig{file=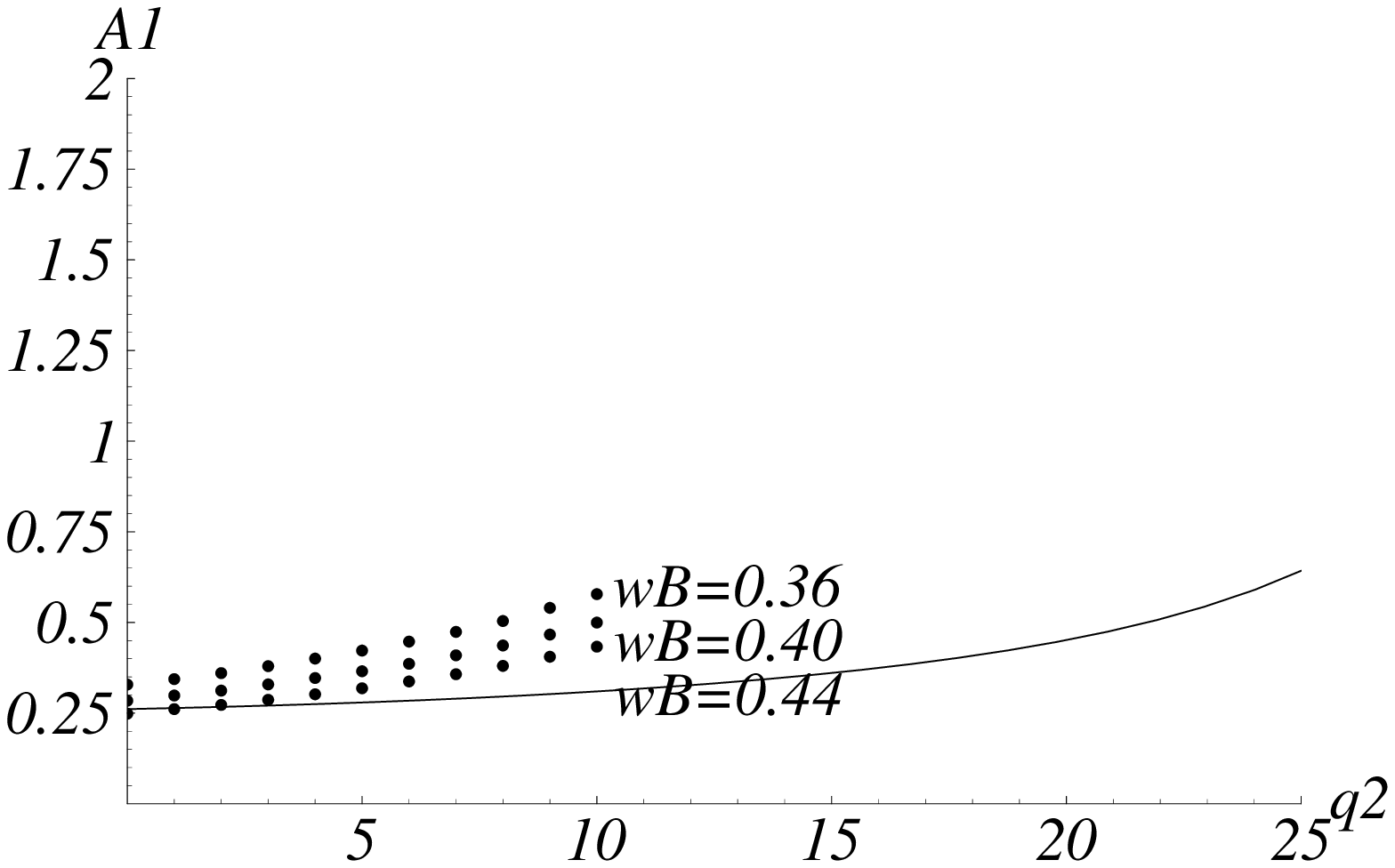,height=5.3cm}
\epsfig{file=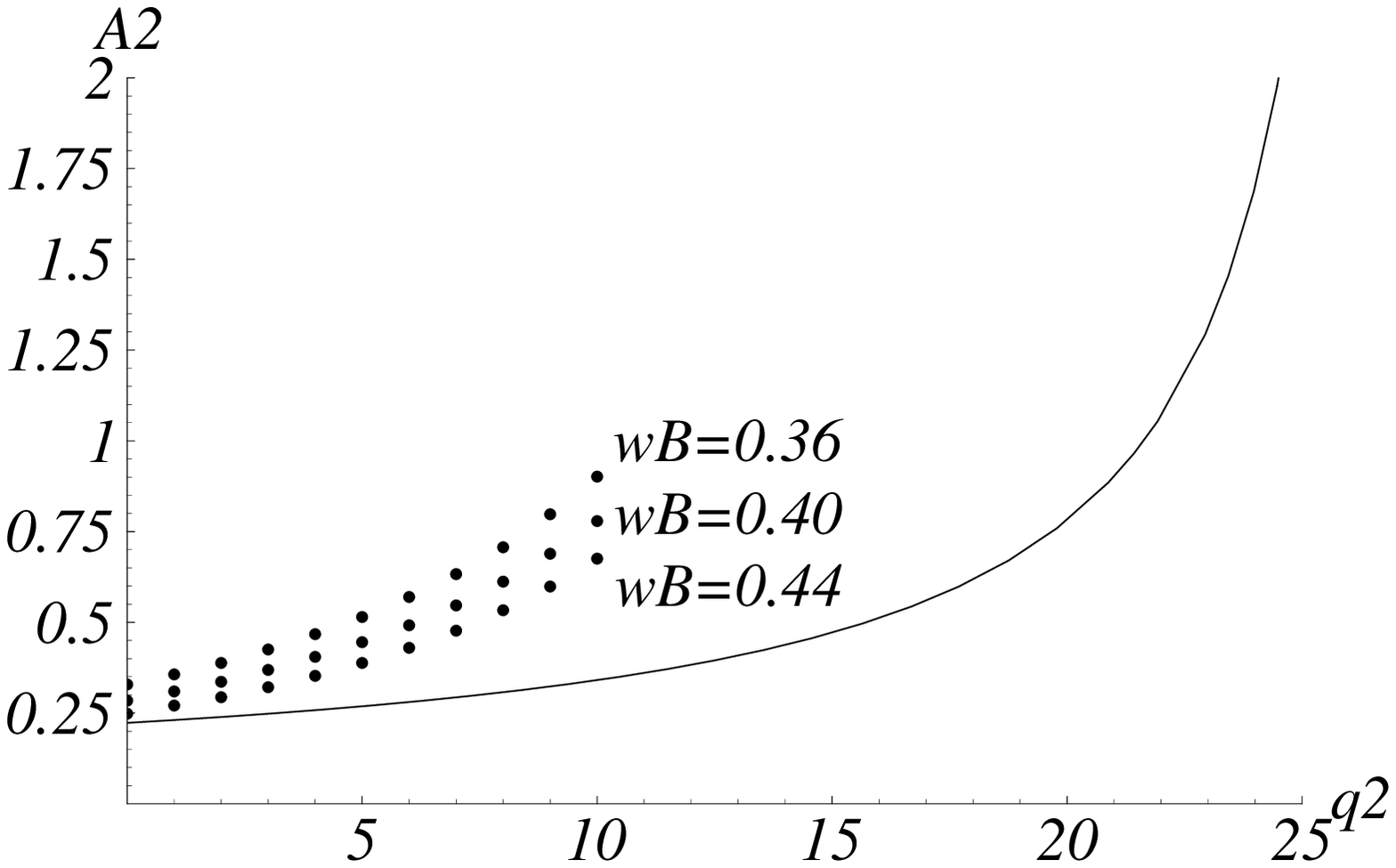,height=5.3cm}
\end{center}
\caption{%
The $B\to \rho$ form factors $V$, $A_0$, $A_1$ and $A_2$ as
functions of $q^2$. PQCD results are given in dots.
The solid lines come from light-cone sum rules.
}
\label{Brho}
\end{figure}

We adopt the $\rho$ meson distribution amplitudes given in the
Appendix B \cite{PB1},
\begin{eqnarray}
\phi_\rho(x)&=&\frac{3f_\rho}{\sqrt{2N_c}} x(1-x)\left[1+
0.18C_2^{3/2}(2x-1)\right]\;,
\label{pwr}\\
\phi_{\rho}^t(x)&=&\frac{f^T_{\rho}}{2\sqrt{2N_c}}
\left\{3(2x-1)^2+0.3(2x-1)^2[5(2x-1)^2-3]\right.
\nonumber \\
& &\left.+0.21[3-30(2x-1)^2+35(2x-1)^4]\right\}\;,
\label{pwt}\\
\phi_{\rho}^s(x) &=&\frac{3f_\rho^T}{2\sqrt{2N_{c}}}
(1-2x)\left[1+0.76(10x^2-10x+1)\right]\;,
\label{pws}\\
\phi_\rho^T(x)&=&\frac{3f_\rho^T}{\sqrt{2N_c}} x(1-x)\left[1+
0.2C_2^{3/2}(2x-1)\right]\;,
\label{pwft}\\
\phi_{\rho}^v(x)&=&\frac{f_{\rho}}{2\sqrt{2N_c}}
\bigg\{\frac{3}{4}[1+(2x-1)^2]+0.24[3(2x-1)^2-1]
\nonumber \\
& &+0.12[3-30(2x-1)^2+35(2x-1)^4]\bigg\}\;,
\label{pwv}\\
\phi_{\rho}^a(x) &=&\frac{3f_\rho}{4\sqrt{2N_{c}}}
(1-2x)\left[1+0.93(10x^2-10x+1)\right]\;,
\label{pwa}
\end{eqnarray}
with the decay constants $f_\rho=200$ MeV and $f_\rho^T=160$ MeV.
The $q^2$ dependence of the form factors $V$ and $A_{0,1,2}$ with the
same $B$ meson distribution amplitude in Eq.~(\ref{os}) and
$M_\rho=0.77$ GeV employed, is displayed in Fig.~\ref{Brho}. Our
results are consistent with those from light-cone QCD sum rules
\cite{sumrho} at small $q^2$.

It is found that the symmetry relation $V=A_1$ in Eq.~(\ref{sr}) holds 
very well: $A_1$ is larger than $V$ only by 2\% in the large recoil 
region, even after considering the pre-asymptotic forms of
$\phi_\rho^v$ and $\phi_\rho^a$ in Eqs.~(\ref{pwv}) and (\ref{pwa}),
respectively. To compare our results with the second symmetry
relation, we include next-to-leading power terms in Eq.~(\ref{ea2}),
obtaining
\begin{eqnarray}
A_2&=&\frac{1+2r_\rho}{\eta}A_1
\nonumber\\
& &-8\pi M_B^2 C_F\frac{2r_\rho}{\eta}
\int dx_1dx_2\int b_1db_1b_2db_2\phi_B(x_1,b_1)
\nonumber\\
& &\times\Biggl\{\left[(1+\eta x_2)\phi_\rho(x_2)
+r_\rho\left(\left(\frac{3}{2\eta}-1\right)\phi_\rho^t(x_2)
+(1-2x_2)\phi_\rho^s(x_2)\right)\right]
E(t^{(1)})h(x_1,x_2,b_1,b_2)
\nonumber \\
& &+2r_\rho\phi_\rho^s(x_2)
E(t^{(2)})h(x_2,x_1,b_2,b_1)\Biggr\}\;.
\end{eqnarray}
Because of the cancellation of the term $2r_\rho A_1/\eta$ and the 
second term in the above expression, the values of $A_2$ only slightly 
deviate from those in Eq.~(\ref{a2f}). The numerical study shows that 
$A_2$ is larger than $A_1-2r_\rho A_0$ by about 40\%, which can be regarded
as the estimate of the symmetry breaking effect.

\section{CONCLUSION}

In this paper we have presented a complete leading-power and leading-order
PQCD evaluation of the $B\to\pi$, $\rho$ transition form factors in the large
recoil region. It has been shown that under Sudakov suppression arising from 
$k_\perp$ and threshold resummations, the end-point singularities (logarithmic 
at twist 2 and linear at twist 3) do not exist. The soft contribution to the 
form factors, being Sudakov suppressed, becomes smaller than the perturbative 
contribution. The physical picture for the mechanism of Sudakov 
suppression has been discussed. We have emphasized that the twist-3 
contributions are in fact not power-suppressed in the $M_B\to \infty$ limit. 
The treatment of the parton transverse momenta $k_\perp$ and the light 
spectator momentum $k_1$ in the $B$ meson in the computation of the hard 
amplitudes has been clearly explained: the hard amplitudes should not be 
expanded in powers of $k_\perp^2$ as the end-point region is important. Using 
the light meson distribution amplitudes derived from QCD sum rules, and choosing 
an appropriate $B$ meson distribution amplitude, we have derived reasonable 
results for the $B\to\pi$, $\rho$ form factors, which are in agreement
with those from light-cone QCD sum rules and from lattice calculations. Our 
study indicates that in a self-consistent perturbative analysis, the 
heavy-to-light form factors are calculable.

The jet function from threshold resummation needs more thorough exploration.
We shall investigate the relevant subjects, such as factorization theorem in
moment space, threshold resummation up to next-to-leading logarithms,
application to nonleptonic $B$ meson decays \cite{CKL}, and numerical effects 
elsewhere. Note that if considering only $k_T$ resummation \cite{DYZ}, twist-3
contributions, though infrared finite, are still too large to give
reasonable heavy-to-light transition form factors, because the large
double logarithms $\alpha_s\ln^2 x$ have not yet been organized.

\vfill
\newpage
\centerline{\large\bf Acknowledgements}

We thank S. Brodsky and H.Y. Cheng for helpful discussions.
The work was supported in part by Grant-in Aid for Special Project
Research (Physics of CP Violation), by Grant-in Aid for Scientific
Research from the Ministry of Education, Science and Culture of Japan.
The work of H.N.L. was supported in part by the National Science Council
of R.O.C. under the Grant No. NSC-89-2112-M-006-033, and by Theory Group
of SLAC. The work of T.K was supported in part by Grant-in Aid for Scientific
Research from the Ministry of Education, Science and Culture of Japan under
the Grant No. 11640265.

\appendix

\section{Pion distribution amplitudes}

It has been shown \cite{L4} that the factorization in fermion flow
between the pion distribution amplitude and the hard amplitude is
achieved by inserting the Fierz identity,
\begin{eqnarray}
I_{ij}I_{lk}=\frac{1}{4}I_{ik}I_{lj}
+\frac{1}{4}(\gamma_5)_{ik}(\gamma_5)_{lj}
+\frac{1}{4}(\gamma_\mu)_{ik}(\gamma^\mu)_{lj}
+\frac{1}{4}(\gamma_5\gamma_\mu)_{ik}(\gamma^\mu\gamma_5)_{lj}
+\frac{1}{8}(\sigma_{\mu\nu})_{ik}(\sigma^{\mu\nu})_{lj}\;,
\label{fe}
\end{eqnarray}
into the quark and anti-quark lines of the pion, where $I$ represents the
identity matrix, and $\sigma^{\mu\nu}$ is defined by
$\sigma^{\mu\nu}=i[\gamma^\mu\gamma^\nu-\gamma^\nu\gamma^\mu]/2$.
The insertion of Eq.~(\ref{fe}) then leads to various nonlocal matrix
elements,
\begin{eqnarray}
\langle 0|{\bar u}(0)\gamma_5\gamma_\mu d(z)|\pi^-(P)\rangle\;,\;\;\;
\langle 0|{\bar u}(0)\gamma_5 d(z)|\pi^-(P)\rangle\;,\;\;\;
\langle 0|{\bar u}(0)\gamma_5\sigma_{\mu\nu}d(z)|\pi^-(P)\rangle\;,\;\;\;
\cdots
\end{eqnarray}
each of which is characterized by different twists. The light-like
vector $z=(0,z^-,0_\perp)$ is the coordinate of the $d$ quark, and
$P=(P^+,0,0_\perp)$ the pion momentum.

The general expressions of the relevant matrix elements are, quoted
from \cite{PB2},
\begin{eqnarray}
\langle 0|{\bar u}(0)\gamma_\mu \gamma_5d(z)|\pi^-(P)\rangle&=&
i\frac{f_\pi}{N_c} P_{\mu}\int_0^1 dx e^{-ix P\cdot z}\phi_v(x)
\nonumber\\
& &+\frac{i}{2}\frac{f_\pi}{N_c} M_\pi^2\frac{z_\mu}{P\cdot z}
\int_0^1 dx e^{-ix P\cdot z}g_\pi(x)\;,
\label{pv}\\
\langle 0|{\bar u}(0)\gamma_5 d(z)|\pi^-(P)\rangle&=&
-i\frac{f_\pi}{N_c} m_{0}\int_0^1 dx e^{-ix P\cdot z}\phi_p(x)\;,
\label{ps}\\
\langle 0|{\bar u}(0)\gamma_5\sigma_{\mu\nu}d(z)|\pi^-(P)\rangle&=&
\frac{i}{6}\frac{f_\pi}{N_c} m_{0}\left(1-\frac{M_\pi^2}{m_{0}^2}\right)
(P_{\mu}z_\nu-P_{\nu}z_\mu)
\int_0^1 dx e^{-ix P\cdot z}\phi_\sigma(x)\;,
\label{pt}
\end{eqnarray}
where $\phi$ and $g_\pi$ are the distribution amplitudes of unit
normalization, $M_\pi$ the pion mass,
$x$ the momentum fraction associated with the $d$ quark.
It is easy to observe that the contribution from $\phi_v$, independent
of the pion mass, is twist-2, and the contribution from $g_\pi$ is twist-4
because of the factor $M_\pi^2$. The contributions from $\phi_p$ and
$\phi_\sigma$, proportional to $r_\pi=m_{0}/M_B$, are twist-3.
We shall neglect the twist-4 terms and
the term $(M_\pi/m_{0})^2$ in Eq.~(\ref{pt}).

It is straightforward to read off the pseudo-vector and pseudo-scalar
structures of the pion distribution amplitudes from Eqs.~(\ref{pv}) and
(\ref{ps}). To derive the pseudo-tensor structure from Eq.~(\ref{pt}),
we need more effort. Using integration by parts, Eq.~(\ref{pt}) is
rewritten as
\begin{eqnarray}
\langle 0|{\bar u}(0)\gamma_5\sigma_{\mu\nu} d(z)|\pi^-(P)\rangle=
\frac{1}{6}\frac{f_\pi}{N_c} m_{0}\left(1-\frac{M_\pi^2}{m_{0}^2}\right)
\epsilon_{\mu\nu}\int_0^1 dx e^{-ix P\cdot z}
\frac{d}{dx}\phi_\sigma(x)\;,
\label{pt3}
\end{eqnarray}
with the anti-symmetric tensor $\epsilon_{\mu\nu}$, $\epsilon^{+-}=1$.
The tensor $\epsilon_{\mu\nu}$ in Eq.~(\ref{pt3}) contracts to the spin
structure $\sigma^{\mu\nu}\gamma_5/2$ in the evaluation of the
corresponding hard amplitude. The factor $1/2$ comes from the extra
factor $1/2$ associated with the pseudo-tensor structure compared to
other structures in Eq.~(\ref{fe}). We have
\begin{eqnarray}
\frac{1}{2}\epsilon_{\mu\nu}\sigma^{\mu\nu}\gamma_5
=-\frac{i}{2}(\gamma^+\gamma^--\gamma^-\gamma^+)\gamma_5
=-i(\not n_-\not n_+-1)\gamma_5\;.
\label{es}
\end{eqnarray}
Therefore, up to twist-3, the initial-state $\pi^-$ meson distribution
amplitudes are written as
\begin{eqnarray}
\langle 0|{\bar u}(0)_jd(z)_l|\pi^-(P)\rangle&=&-\frac{i}{\sqrt{2N_c}}
\int_0^1dx e^{-ixP\cdot z}\left\{[\not P\gamma_5]_{lj}\phi_\pi(x)
+ [\gamma_5]_{lj}m_0\phi^p_\pi(x)\right.
\nonumber\\
& &\left.+m_0[\gamma_5(\not n_-\not n_+-1)]_{lj}\phi^t_\pi(x)\right\}\;,
\end{eqnarray}
with
\begin{eqnarray}
\phi_\pi(x)=\frac{f_\pi}{2\sqrt{2N_c}}\phi_{v}(x)\;,\;\;\;
\phi_\pi^p(x)=\frac{f_\pi}{2\sqrt{2N_c}}\phi_p(x)\;,\;\;\;
\phi_\pi^t(x)=\frac{f_\pi}{12\sqrt{2N_c}}
\frac{d}{dx}\phi_\sigma(x)\;.
\end{eqnarray}

For the final-state $\pi^-$ meson, we consider the adjoints of
Eqs.~(\ref{pv}), (\ref{ps}) and (\ref{pt}):
\begin{eqnarray}
\langle \pi^-(P)|{\bar d}(z)\gamma_\mu\gamma_5u(0)|0\rangle&=&
-i\frac{f_\pi}{N_c} P_\mu\int_0^1 dx e^{ix P\cdot z}\phi_v(x)\;,
\label{pvo}\\
\langle \pi^-(P)|{\bar d}(z)\gamma_5 u(0)|0\rangle&=&
-i\frac{f_\pi}{N_c} m_{0}\int_0^1 dx e^{ix P\cdot z}\phi_p(x)\;,
\label{ps0}\\
\langle \pi^-(P^-)|{\bar d}(z)\sigma_{\mu\nu}\gamma_5 u(0)|0\rangle&=&
-\frac{f_\pi}{6N_c} m_{0}\left(1-\frac{M_\pi^2}{m_{0}^2}\right)
\epsilon_{\mu\nu}\int_0^1 dx e^{ix P\cdot z}
\frac{d}{dx}\phi_\sigma(x)\;.
\label{pto}
\end{eqnarray}
It is observed that the pseudo-tensor structure in Eq.~(\ref{pto})
acquires an extra minus sign, compared to the other two structures.
The pseudo-tensor structure is then given by
$-\gamma_5(\not n_-\not n_+-1)=\gamma_5(\not n_+\not n_--1)$.
Therefore, up to twist 3, we have Eq.~(\ref{fpd}) for the final-state
$\pi^-$ meson. Note that there is an extra term in the definition of
$\phi_\pi^t$, which contains a differential operator applying to hard
amplitudes \cite{BF}. This term, being power-suppressed, is negligible
here. The distribution amplitudes $\phi_\pi$ and $\phi_\pi^p$ are normalized
according to
\begin{eqnarray}
\int_0^1 dx\phi_\pi(x)=\frac{f_\pi}{2\sqrt{2N_c}}\;,\;\;\;
\int_0^1 dx\phi_\pi^p(x)=\frac{f_\pi}{2\sqrt{2N_c}}\;.
\end{eqnarray}
The tensor distribution amplitude is normalized to zero, because of
\begin{eqnarray}
\int_0^1 dx \frac{d}{dx}\phi_{\sigma}(x)
=\phi_{\sigma}(1)-\phi_{\sigma}(0)=0\;,
\label{zen}
\end{eqnarray}
if $\phi_\sigma$ vanishes at the end points of the momentum fraction.

\section{$\rho$ Meson distribution amplitudes}

We choose the $\rho$ meson momentum $P$ with $P^2=M_\rho^2$, which is
mainly in the plus direction. The polarization vectors $\epsilon$,
satisfying $P\cdot \epsilon=0$, represent one longitudinal polarization
vector $\epsilon_L$ and two transverse polarization vectors $\epsilon_T$.
Their explicit expressions in light-cone coordinates have been given
in Eq.~(\ref{par}).
To arrive at the factorization in fermion flow, we insert the Fierz
identity into the quark and anti-quark lines of the
$\rho$ meson. The spin structures in Eq.~(\ref{fe}) lead to the following
nonlocal matrix elements,
\begin{eqnarray}
& &\langle \rho^-(P,\epsilon)|{\bar d}(z)\gamma_\mu u(0)|0
\rangle\;,\;\;\;
\langle \rho^-(P,\epsilon)|{\bar d}(z)\sigma_{\mu\nu}u(0)|0
\rangle\;,
\nonumber\\
& &\langle \rho^-(P,\epsilon)|{\bar d}(z)I u(0)|0\rangle\;,\;\;\;
\langle \rho^-(P,\epsilon)|{\bar d}(z)\gamma_\mu\gamma_5 u(0)|0
\rangle\;,
\end{eqnarray}
characterized by different twists. The definition of $z$ is the same as
that for the pion distribution amplitudes in the previous Appendix.

The general expressions of the above matrix elements are, quoted
from \cite{PB1},
\begin{eqnarray}
\langle \rho^-(P,\epsilon)|{\bar d}(z)\gamma_\mu u(0)|0\rangle&=&
\frac{f_{\rho}}{N_c}M_{\rho}\left\{P_{\mu}\frac{\epsilon\cdot z}{P\cdot z}
\int_0^1 dx e^{ix P\cdot z}\phi_{\parallel}(x)
+\epsilon_{T\mu}\int_0^1 dx e^{ix P\cdot z}g^{(v)}_T(x)\right.
\nonumber\\
& &\left. -\frac{1}{2}z_\mu\frac{\epsilon\cdot z}{(P\cdot z)^2}
M^2_{\rho}\int_0^1 dx e^{ix P\cdot z}[g_3(x)-\phi_\parallel(x)]\right\}\;,
\label{v}\\
\langle \rho^-(P,\epsilon)|{\bar d}(z)\sigma_{\mu\nu}u(0)|0\rangle&=&
-i\frac{f^T_{\rho}}{N_c}\left\{(\epsilon_{T\mu}P_{\nu}-\epsilon_{T\nu}P_{\mu})
\int_0^1 dx e^{ix P\cdot z}\phi_T(x)\right.
\nonumber\\
& &+(P_{\mu} z_\nu-P_{\nu} z_\mu)\frac{\epsilon\cdot z}{(P\cdot z)^2}
M^2_{\rho}\int_0^1 dx e^{ix P\cdot z}h^{(t)}_{\parallel}(x)
\nonumber\\
& &\left. +\frac{1}{2}(\epsilon_{T\mu}z_\nu-\epsilon_{T\nu}z_{\mu})
\frac{M^2_{\rho}}{P\cdot z}\int_0^1 dx e^{ix P\cdot z}[h_3(x)-\phi_T(x)]
\right\}\;,
\label{t}\\
\langle \rho^-(P,\epsilon)|{\bar d}(z)I u(0)|0\rangle&=&
-\frac{i}{2N_c}\left(f^T_{\rho}-f_{\rho}\frac{m_u+m_d}{M_{\rho}}\right)
\epsilon\cdot zM^2_{\rho}\int_0^1 dx e^{ix P\cdot z}
h^{(s)}_{\parallel}(x)\;,
\nonumber\\
&=&\frac{1}{2N_c}\left(f^T_{\rho}-f_{\rho}\frac{m_u+m_d}{M_{\rho}}\right)
\frac{\epsilon\cdot z}{P\cdot z}M^2_{\rho}
\int_0^1 dx e^{ix P\cdot z}
\frac{d}{dx}h^{(s)}_{\parallel}(x)\;,
\label{s}\\
\langle \rho^-(P,\epsilon)|{\bar d}(z)\gamma_5\gamma_\mu u(0)|0\rangle
&=&-\frac{1}{4N_c}\left(f_{\rho}-f_{\rho}^T\frac{m_u+m_d}{M_{\rho}}\right)
M_\rho\epsilon_{\mu}^{~\nu\alpha\beta}\epsilon_{T\nu} P_{2\alpha}
z_\beta\int_0^1 dx e^{ix P\cdot z}g^{(a)}_{T}(x)\;,
\nonumber\\
&=&-\frac{i}{4N_c}\left(f_{\rho}-f_{\rho}^T\frac{m_u+m_d}{M_{\rho}}\right)
\frac{M_\rho}{P\cdot n_-}\epsilon_{\mu\nu\rho\sigma}\epsilon_T^\nu
P^\rho n_-^\sigma\int_0^1 dx e^{ix P\cdot z}
\frac{d}{dx}g^{(a)}_{T}(x)\;,
\label{a}
\end{eqnarray}
where $f_{\rho}$ and $f^T_{\rho}$ are the decay constants of the $\rho$
meson with longitudinal and transverse polarizations, respectively, and
$x$ the momentum fraction associated with the $d$ quark. We adopt the
convention $\epsilon^{0123}=1$ for the Levi-Civita tensor
$\epsilon^{\mu\nu\alpha\beta}$. The distribution amplitudes $\phi$,
$g$ and $h$ are normalized to unity.

Following the similar procedures, we derive the $\rho$ meson distribution
amplitudes up to twist 3,
\begin{eqnarray}
\langle \rho^-(P,\epsilon_L)|\bar d(z)_ju(0)_l|0\rangle
&=&\frac{1}{\sqrt{2N_c}}\int_0^1 dx e^{ixP\cdot z}
\left\{M_\rho[\not \epsilon_L]_{lj}\phi_\rho(x)\right.
\nonumber\\
& &\left.+[\not \epsilon_L\not P]_{lj} \phi_{\rho}^{t}(x)
+M_\rho [I]_{lj}\phi_\rho^s(x)\right\}\;,
\label{lpf}\\
\langle\rho^-(P,\epsilon_T)|\bar d(z)_ju(0)_l|0\rangle
&=&\frac{1}{\sqrt{2N_c}}\int_0^1 dx e^{ixP\cdot z}
\bigg\{M_\rho[\not \epsilon_T]_{lj}\phi_\rho^v(x)+
[\not\epsilon_T\not P]_{lj}\phi_\rho^T(x)
\nonumber\\
& &+\frac{M_\rho}{P\cdot n_-}
i\epsilon_{\mu\nu\rho\sigma}[\gamma_5\gamma^\mu]_{lj}\epsilon_T^\nu
P^\rho n_-^\sigma \phi_\rho^a(x)\bigg\}\;,
\label{spf}
\end{eqnarray}
for longitudinal polarization and transverse polarization, respectively.
We have dropped the terms proportional to $r_\rho^2$ (twist-4) and the
terms $(m_u+m_d)/M_{\rho}$ in Eqs.~(\ref{s}) and (\ref{a}).
The definitions of the above distribution amplitudes are,
\begin{eqnarray}
& &\phi_{\rho}=\frac{f_{\rho}}{2\sqrt{2N_c}}\phi_{\parallel}\;,\;\;\;
\phi_{\rho}^t=\frac{f^T_{\rho}}{2\sqrt{2N_c}}h^{(t)}_{\parallel}\;,\;\;\;
\phi_{\rho}^s=\frac{f^T_{\rho}}{4\sqrt{2N_c}}
\frac{d}{dx}h^{(s)}_{\parallel}\;,
\\
&&\phi_{\rho}^T=\frac{f_{\rho}^T}{2\sqrt{2N_c}}\phi_T\;,\;\;\;
\phi_{\rho}^v=\frac{f_{\rho}}{2\sqrt{2N_c}}g^{(v)}_T\;,\;\;\;
\phi_{\rho}^a=\frac{f_{\rho}}{8\sqrt{2N_c}}
\frac{d}{dx}g_T^{(a)}\;.
\end{eqnarray}

\section{$B$ Meson distribution amplitudes}

According to \cite{GN}, the nonlocal matrix element associated
with the $B$ meson is written as
\begin{eqnarray}
&&\int \frac{d^4z}{(2\pi)^4}e^{ik_1\cdot z}
\langle 0|{\bar b}_{\alpha}(0)d_{\delta}(z)|B(P_1)\rangle
\nonumber\\
&=&\frac{i}{\sqrt{2N_c}}\left\{(\not P_1+M_B)\gamma_5
\left[\frac{\not n_+}{\sqrt{2}}\phi_B^{+}(k_1)
+\frac{\not n_-}{\sqrt{2}}\phi_B^{-}(k_1)\right]
\right\}_{\delta\alpha}\;,
\nonumber\\
&=&-\frac{i}{\sqrt{2N_c}}\left\{(\not P_1+M_B)\gamma_5
\left[\phi_B(k_1)-\frac{\not n_+-\not n_-}{\sqrt{2}}
{\bar \phi}_B(k_1)\right]\right\}_{\delta\alpha}\;,
\label{bwp2}
\end{eqnarray}
with the wave functions,
\begin{eqnarray}
\phi_B=\frac{1}{2}(\phi_B^{+}+\phi_B^{-})\;,\;\;\;
{\bar\phi}_B=\frac{1}{2}(\phi_B^{+}-\phi_B^{-})\;.
\end{eqnarray}
Because the light meson momenta have been chosen in the plus direction, the
hard amplitudes for the heavy-to-light transition form factors are
independent of the component $k_{1}^+$ as explained in
Sec.~III. We construct the $B$ meson distribution amplitude,
\begin{eqnarray}
\phi(x_{1},b)=\int dk_{1}^+d^2k_{1\perp} e^{i{\vec k}_{1\perp}\cdot {\vec b}}
\phi(k_{1})\;,
\end{eqnarray}
with $x_1=k_1^-/P_1^-$.

The two $B$ meson distribution amplitudes $\phi_B^{+}(x)=\phi_B^+(x,0)$
and $\phi_B^{-}(x)=\phi_B^-(x,0)$ are related by the equation of motion
\cite{GN},
\begin{eqnarray}
\phi_B^+(x)=-x\frac{d}{dx}\phi_B^{-}(x)\;.
\label{beq}
\end{eqnarray}
Assuming that $\phi_B^-$ vanishes at both ends of the momentum fraction,
$x\to 0$ and $x\to 1$, we derive
\begin{eqnarray}
& &\int_0^1dx\phi_B^+(x)=\int_0^1dx\phi_B^-(x)
=\frac{f_B}{2\sqrt{2N_c}}\;,
\nonumber\\
& &\int_0^1dx x\phi_B^+(x)=2\int_0^1dx x\phi_B^-(x)\sim
\frac{2\bar\Lambda}{M_B}\frac{f_B}{2\sqrt{2N_c}}\;.
\label{bep}
\end{eqnarray}
Therefore, $\bar\phi_B$ is normalized to zero.

We shall argue that the contribution from the distribution amplitude
$\bar\phi_B$ is negligible compared to that from $\phi_B$.
Consider the reasonable parametrizations,
\begin{eqnarray}
\phi_B(x)&=&\frac{f_B}{2\sqrt{2N_c}}\left[
\delta\left(x-\frac{\bar\Lambda}{M_B}\right)
-\frac{\bar\Lambda}{2M_B}\delta'\left(x-\frac{\bar\Lambda}{M_B}\right)
+O\left(\frac{\bar\Lambda^2}{M_B^2}\right)\right]\;,
\nonumber\\
\bar\phi_B(x)&=&\frac{f_B}{2\sqrt{2N_c}}
\left[-\frac{\bar\Lambda}{2M_B}\delta'\left(x-\frac{\bar\Lambda}
{M_B}\right)+O\left(\frac{\bar\Lambda^2}{M_B^2}\right)\right]\;,
\end{eqnarray}
whose moments satisfy Eq.~(\ref{bep}).
As shown in Sec.~III, the hard amplitudes are approximated by $\ln(1/x_1)$ 
at small $x_1$. A simple estimation indicates that the contribution from
$\phi_B$, proportional to $\ln(M_B/\bar\Lambda)$, is numerically larger than 
that from $\bar\phi_B$, proportional to a constant. Hence, after taking
into account Eq.~(\ref{beq}), we consider only a single $B$ meson
distribution amplitude in this work.

\section{Threshold resummation\label{sudd}}

In this Appendix we supply details of the derivation of the Sudakov factor in
Eq.~(\ref{trs}). Threshold resummation introduces a jet function $S_t(x)$
into the PQCD factorization of the $B\to\pi$ form factors near the
end points \cite{L3},
\begin{eqnarray}
S_t(x)=\int_{a-i\infty}^{a+i\infty}\frac{dN}{2\pi i}
\frac{S_t(N)}{N}(1-x)^{-N}\;,
\label{mj}
\end{eqnarray}
where $a$ is an arbitrary real constant larger than all the real
parts of poles involved in the integrand.
The factor $1/N$ comes from Mellin transformation of the
initial condition $S^{(0)}_t(x)=1$,
\begin{eqnarray}
\int_0^1 dx (1-x)^{N-1}S^{(0)}_t(x)=\frac{1}{N}\;.
\end{eqnarray}
The Sudakov factor $S_t(N)$ in the moment $(N)$ space has been derived
explicitly to the accuracy of leading logarithms (LL) \cite{L3},
\begin{eqnarray}
S_t^{(LL)}(N)=\exp\left[-\frac{1}{4}\gamma_K^{(LL)}\ln^2 N\right]\;,
\end{eqnarray}
with the anomalous dimension $\gamma_K^{(LL)}=\alpha_sC_F/\pi$.
The contour integral in Eq.~(\ref{mj}) leads to
\begin{eqnarray}
S_t^{(LL)}(x)=-\exp\left(\frac{\pi}{4}\alpha_sC_F\right)
\int_{-\infty}^{\infty}\frac{dt}{\pi}(1-x)^{\exp(t)}
\sin\left(\frac{1}{2}\alpha_s C_Ft\right)
\exp\left(-\frac{\alpha_s}{4\pi}C_Ft^2\right)\;.
\label{mjx}
\end{eqnarray}
which vanishes at $x\to 0$, since the integrand
is an odd function in $t$, and at $x\to 1$ due to the factor
$(1-x)^{\exp(t)}$.

In this paper we consider threshold resummation up to next-to-leading
logarithms. At this level of accuracy, the anomalous dimension
$\gamma_K$ contains two-loop contributions, and the coupling
constant $\alpha_s$ becomes running.
The Sudakov factor $S_t(N)$ is then given by
\begin{eqnarray}
S_t(N)=\exp\left[\frac{1}{2}\int_0^{1-1/N}\frac{dz}{1-z}
\int_{(1-z)}^{(1-z)^2}
\frac{d\lambda}{\lambda}\gamma_K(\alpha_s(\lambda M_B^2/2))\right]\;,
\label{exa}
\end{eqnarray}
with 
\begin{equation}
\gamma_K=\frac{\alpha_s}{\pi}C_F+\left(\frac{\alpha_s}{\pi}
\right)^2C_F\left[C_A\left(\frac{67}{36}
-\frac{\pi^{2}}{12}\right)-\frac{5}{18}n_{f}\right]\;,
\label{lk}
\end{equation}
$n_{f}$ being the number of quark flavors and $C_A=3$ a color factor.
The anomalous dimension $\gamma_K$ is the same as that for $k_T$ 
resummation \cite{L2}.

\begin{figure}[h]
\begin{center}
\epsfig{file=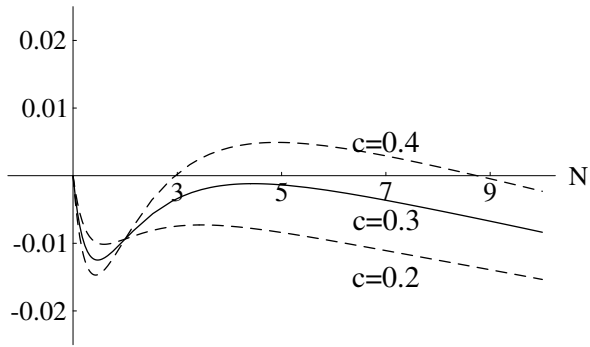,width=8cm}
\end{center}
\caption{Difference between the jet function and its parametrization
in the moment space.
}
\label{thfit}
\end{figure}

It can be shown that $S_t(x)$ still vanishes at the end points
$x\to 0$ and $x\to 1$. To simplify the analysis, we propose the
parametrization,
\begin{eqnarray}
S_t(x)=\frac{2^{1+2c}\Gamma(3/2+c)}{\sqrt{\pi}\Gamma(1+c)} [x(1-x)]^c\;,
\label{pst}
\end{eqnarray}
whose end-point behavior satisfies the above requirement. In the 
$\alpha_s\to 0$ ($c\to 0$) limit, {\it i.e.}, without QCD effects,
Eq.~(\ref{pst}) approaches unity. Mellin transformation of $S_t(x)$ 
gives
\begin{equation}
\frac{S_t^{fit}(N)}{N}\equiv
\frac{2^{1+2c}\Gamma(3/2+c)}{\sqrt{\pi}\Gamma(1+c)} B(c+1,c+N)\;.
\label{trf}
\end{equation}
The variable $N$ should be large enough to justify
threshold resummation up to the next-to-leading logarithms
$\alpha_s\ln N$, and small enough to avoid the divergent
running coupling constant $\alpha_s(M_B^2/(2N^2))$ in Eq.~(\ref{exa}).
Performing the best fit of Eq.~(\ref{trf}) to $S_t(N)/N$ for
$3<N<7$, we determine the parameter $c=0.3$. The difference
$S_t(N)/N - S_t^{fit}(N)/N$ is shown in Fig.~\ref{thfit}
for $c = 0.2,0.3$ and 0.4. Equation (\ref{pst}) implies that threshold 
resummation modifies the end-point behavior of the meson distribution 
amplitudes, rendering them vanish faster at $x\to 0$.

\end{document}